\documentclass[12pt]{article}
\pdfoutput=1
\usepackage[colorlinks,linkcolor=Blue,citecolor=Blue,bookmarks,bookmarksnumbered]{hyperref}
\usepackage[scaled=0.85]{helvet}
\usepackage{amsmath,amssymb,accents,mathrsfs,XoohmE}
\usepackage{graphicx,color}
\graphicspath{{Figures/}}
\usepackage{booktabs}
\usepackage{multirow}
\usepackage{placeins}
\usepackage{amsmath}
\usepackage{subfigure}

\usepackage{XoohmE}

\definecolor{Green}  {rgb}{0.10,0.70,0.10} 
\definecolor{Orange} {rgb}{1.00,0.50,0.15} 
\definecolor{Red}    {rgb}{0.90,0.00,0.12} 
\definecolor{Purple} {rgb}{0.50,0.25,0.55} 
\definecolor{Turque} {rgb}{0.00,0.65,0.85} 
\definecolor{Blue}   {rgb}{0.00,0.00,1.00} 
\definecolor{Magenta}{rgb}{1.00,0.00,1.00} 
\definecolor{Gold}   {rgb}{1.00,0.75,0.25} 
\definecolor{Seaweed}{rgb}{0.01,0.24,0.09} 
\definecolor{Brown}  {rgb}{0.43,0.26,0.32} 
\definecolor{grey1}  {rgb}{0.20,0.20,0.20} 
\definecolor{grey2}  {rgb}{0.40,0.40,0.40} 
\definecolor{grey3}  {rgb}{0.60,0.60,0.60} 
\definecolor{grey4}  {rgb}{0.80,0.80,0.80} 
\definecolor{grey5}  {rgb}{0.90,0.90,0.90} 
\def\C#1#2{{\ifcase#1\or
             \color{Green}\or \color{Orange}\or \color{Red}\or
              \color{Purple}\or \color{Turque}\or \color{Blue}\or
               \color{Magenta}\or \color{Gold}\or \color{Seaweed}\or
                \color{Brown}\or\color{grey1}\or\color{grey2}\or
                 \color{grey3}\else\color{grey4}\fi#2}}

\definecolor{Slate} {rgb}{0.00,0.45,0.55}

\def\fracm#1#2{\hbox{\large{${\frac{{#1}}{{#2}}}$}}}

\def\be{\begin{equation}}
\def\ee{\end{equation}}
\newcommand{\bea}{\begin{eqnarray}}
\newcommand{\eea}{\end{eqnarray}}
\newcommand{\ena}{\end{eqnarray}}

\def\pp{{\mathchoice
          {
              \kern 1pt%
              \raise 1pt
              \vbox{\hrule width5pt height0.4pt depth0pt
                    \kern -2pt
                    \hbox{\kern 2.3pt
                          \vrule width0.4pt height6pt depth0pt
                          }
                    \kern -2pt
                    \hrule width5pt height0.4pt depth0pt}%
                    \kern 1pt
           }
            {
              \kern 1pt%
              \raise 1pt
              \vbox{\hrule width4.3pt height0.4pt depth0pt
                    \kern -1.8pt
                    \hbox{\kern 1.95pt
                          \vrule width0.4pt height5.4pt depth0pt
                          }
                    \kern -1.8pt
                    \hrule width4.3pt height0.4pt depth0pt}%
                    \kern 1pt
            }
            {
              \kern 0.5pt%
              \raise 1pt
              \vbox{\hrule width4.0pt height0.3pt depth0pt
                    \kern -1.9pt  
                    \hbox{\kern 1.85pt
                          \vrule width0.3pt height5.7pt depth0pt
                          }
                    \kern -1.9pt
                    \hrule width4.0pt height0.3pt depth0pt}%
                    \kern 0.5pt
            }
            {
              \kern 0.5pt%
              \raise 1pt
              \vbox{\hrule width3.6pt height0.3pt depth0pt
                    \kern -1.5pt
                    \hbox{\kern 1.65pt
                          \vrule width0.3pt height4.5pt depth0pt
                          }
                    \kern -1.5pt
                    \hrule width3.6pt height0.3pt depth0pt}%
                    \kern 0.5pt
            }
        }}

\def\mm{{\mathchoice
                       {
                             \kern 1pt
               \raise 1pt    \vbox{\hrule width5pt height0.4pt depth0pt
                                  \kern 2pt
                                  \hrule width5pt height0.4pt depth0pt}
                             \kern 1pt}
                       {
                            \kern 1pt
               \raise 1pt \vbox{\hrule width4.3pt height0.4pt depth0pt
                                  \kern 1.8pt
                                  \hrule width4.3pt height0.4pt depth0pt}
                             \kern 1pt}
                       {
                            \kern 0.5pt
               \raise 1pt
                            \vbox{\hrule width4.0pt height0.3pt depth0pt
                                  \kern 1.9pt
                                  \hrule width4.0pt height0.3pt depth0pt}
                            \kern 1pt}
                       {
                           \kern 0.5pt
             \raise 1pt  \vbox{\hrule width3.6pt height0.3pt depth0pt
                                  \kern 1.5pt
                                  \hrule width3.6pt height0.3pt depth0pt}
                           \kern 0.5pt}
                       }}

\def\ad{{\kern0.5pt
                   \alpha \kern-5.05pt \raise5.8pt\hbox{$\textstyle.$}\kern
0.5pt}}

\def\bd{{\kern0.5pt
                   \beta \kern-5.05pt \raise5.8pt\hbox{$\textstyle.$}\kern
0.5pt}}

\def\qd{{\kern0.5pt
                   q \kern-5.05pt \raise5.8pt\hbox{$\textstyle.$}\kern
0.5pt}}
\def\Dot#1{{\kern0.5pt
     {#1} \kern-5.05pt \raise5.8pt\hbox{$\textstyle.$}\kern
0.5pt}}

\catcode`@=11
\def\un#1{\relax\ifmmode\@@underline#1\else
        $\@@underline{\hbox{#1}}$\relax\fi}
\catcode`@=12

\def\a{\alpha}
\def\b{\beta}
\def\c{\chi}
\def\d{\delta}
\def\e{\epsilon}

\def\g{\gamma}

\def\i{\iota}
\def\j{\psi}
\def\k{\kappa}
\def\l{\lambda}
\def\m{\mu}
\def\n{\nu}
\def\o{\omega}

\def\r{\rho}
\def\s{\sigma}
\def\t{\tau}

\def\L{\Lambda}
\def\O{\Omega}
\def\P{\Pi}

\def\S{\Sigma}

\def\ca{{\cal A}}

\def\cf{{\cal F}}

\def\dslash{\not{\hbox{\kern-2pt $\partial$}}}
\def\Dslash{\not{\hbox{\kern-4pt $D$}}}
\def\pslash{\not{\hbox{\kern-2.3pt $p$}}}
 \newtoks\slashfraction
 \slashfraction={.13}
 \def\slash#1{\setbox0\hbox{$ #1 $}
 \setbox0\hbox to \the\slashfraction\wd0{\hss \box0}/\box0 }
 
\def\kcr{{\hbox{\ro \char'170}}}                
\def\ktl{{\hbox{\ro \char'170}}}        
\def\ktr{{\hbox{\ro \char'170}}}        
\def\kbl{{\hbox{\ro \char'170}}}        
\def\kbr{{\hbox{\ro \char'170}}}        

\def\plpl{\raise-2pt\hbox{$\raise3pt\hbox{$_+$}\hskip-6.67pt\raise0.0pt
\hbox{$^+$}\hskip 0.01pt$}}
\def\mimi{\raise-2pt\hbox{$\raise3pt\hbox{$_-$}\hskip-6.67pt\raise0.0pt
\hbox{$^-$}\hskip 0.01pt$}} 

\def\bo{{\raise.15ex\hbox{\large$\Box$}}}               
\def\pa{\partial}                                       
\def\TH{{\raise.2ex\hbox{$\displaystyle \bigodot$}\mskip-4.7mu \llap H \;}}
\def\face{{\raise.2ex\hbox{$\displaystyle \bigodot$}\mskip-2.2mu \llap {$\ddot
        \smile$}}}                                      

\def\dt#1{\on{\hbox{\bf .}}{#1}}                
\def\Dot#1{\dt{#1}}

\def\Tilde#1{\widetilde{#1}}                    
\def\Hat#1{\widehat{#1}}                        
\def\leftrightarrowfill{$\mathsurround=0pt \mathord\leftarrow \mkern-6mu
        \cleaders\hbox{$\mkern-2mu \mathord- \mkern-2mu$}\hfill
        \mkern-6mu \mathord\rightarrow$}
\def\dvec#1{\vbox{\ialign{##\crcr
        \leftrightarrowfill\crcr\noalign{\kern-1pt\nointerlineskip}
        $\hfil\displaystyle{#1}\hfil$\crcr}}}           
\def\dt#1{{\buildrel {\hbox{\LARGE .}} \over {#1}}}     

\def\fracm#1#2{\hbox{\large{${\frac{{#1}}{{#2}}}$}}}
\def\sfrac#1#2{{\vphantom1\smash{\lower.5ex\hbox{\small$#1$}}\over
        \vphantom1\smash{\raise.4ex\hbox{\small$#2$}}}} 
\def\bfrac#1#2{{\vphantom1\smash{\lower.5ex\hbox{$#1$}}\over
        \vphantom1\smash{\raise.3ex\hbox{$#2$}}}}       
\def\afrac#1#2{{\vphantom1\smash{\lower.5ex\hbox{$#1$}}\over#2}}    

\def\pa{\partial}      
\let\bm\relax
\newcommand{\bm}[1]{{\boldsymbol{#1}}}

\def\ad{{\dot{\alpha}}}
\def\bd{{\dot{\beta}}}

 \font\rOpe=cmsy10                        
 \def\ktl{{\hbox{\rOpe\char'170}}}        
 \def\kbl{{\hbox{\rOpe\char'170}}}        
 \def\kcr{{\reflectbox{\rOpe\char'170}}}        
 \def\ktr{{\reflectbox{\rOpe\char'170}}}        
 \def\kbr{{\reflectbox{\rOpe\char'170}}}        
 \def\Border{\vbox{\hsize0pt
        \setlength{\unitlength}{1mm}
        \newcount\xco
        \newcount\yco
        \xco=-21
        \yco=12
        \begin{picture}(0,0)(-7.5,0)
        \put(\xco,\yco){$\ktl$}
        \advance\yco by-1
        {\loop
        \put(\xco,\yco){$\kcr$}
        \advance\yco by-2
        \ifnum\yco>-240
        \repeat
        \put(\xco,\yco){$\kbl$}}
        \xco=170
        \yco=12
        \put(\xco,\yco){$\ktr$}
        \advance\yco by-1
        {\loop
        \put(\xco,\yco){$\kcr$}
        \advance\yco by-2
        \ifnum\yco>-240
        \repeat
        \put(\xco,\yco){$\kbr$}}
        \put(-19.5,13){\scalebox{.6065}{%
         University of Maryland Center for String and Particle  Theory \&\ Physics Department%
        |University of Maryland Center for String and Particle  Theory \&\ Physics Department}}
        \put(-19.5,-241.5){\scalebox{.5835}{%
         ****University of Maryland * Center for String and
         Particle  Theory* Physics Department****University of Maryland *Center
        for String and Particle  Theory* Physics Department}}
        \end{picture}
        \par\vskip-8mm}}
\definecolor{UMred}{rgb}{.9,.05,.2}
\definecolor{HUblue}{rgb}{.0,.3,.7}

\definecolor{Red}    {rgb}{0.90,0.00,0.12} 
\definecolor{Blue}   {rgb}{0.00,0.00,1.00} 
\definecolor{Green}  {rgb}{0.10,0.70,0.10} 
\definecolor{Turque} {rgb}{0.00,0.65,0.85} 
\definecolor{Orange} {rgb}{1.00,0.50,0.15} 
\definecolor{Magenta}{rgb}{1.00,0.00,1.00} 
\definecolor{Gold}   {rgb}{1.00,0.75,0.25} 
\definecolor{Seaweed}{rgb}{0.01,0.24,0.09} 
\definecolor{Purple} {rgb}{0.50,0.25,0.55} 
\definecolor{Brown}  {rgb}{0.43,0.26,0.32} 
\definecolor{grey1}  {rgb}{0.20,0.20,0.20} 
\definecolor{grey2}  {rgb}{0.40,0.40,0.40} 
\definecolor{grey3}  {rgb}{0.60,0.60,0.60} 
\definecolor{grey4}  {rgb}{0.80,0.80,0.80} 
\definecolor{grey5}  {rgb}{0.90,0.90,0.90} 
\def\C#1#2{{\ifcase#1\or
             \color{Red}\or \color{Green}\or \color{Blue}\or\
              \color{Turque}\or \color{Orange}\or \color{Magenta}\or 
               \color{Gold}\or \color{Seaweed}\or \color{Purple}\or
                \color{Brown}\or\color{grey1}\or\color{grey2}\or
                 \color{grey3}\else\color{grey4}\fi#2}}

\definecolor{Slate} {rgb}{0.00,0.45,0.55}

\newdimen\parshift\parshift=\parindent
\catcode`@=11
 \long\def\@footnotetext#1{\insert\footins{\reset@font\footnotesize
           \interlinepenalty\interfootnotelinepenalty\splittopskip%
            \footnotesep\splitmaxdepth\dp\strutbox\floatingpenalty\@MM%
             \hsize\columnwidth\addtolength{\hsize}{-2\parindent}
              \@parboxrestore\protected@edef\@currentlabel%
              {\csname p@footnote\endcsname\@thefnmark}%
                \color@begingroup%
                 \@makefntext{\rule\z@\footnotesep\ignorespaces#1%
                  \@finalstrut\strutbox}%
                \color@endgroup}}
 \long\def\@makefntext#1{\hglue\parshift%
           \vbox{\noindent\baselineskip=11pt plus.5pt minus.5pt\hb@xt@0em{\hss\@makefnmark\kern1pt}#1}}
\catcode`@=12

\newskip\humongous \humongous=0pt plus 1000pt minus 1000pt
\def\caja{\mathsurround=0pt}
\def\eqalign#1{\,\vcenter{\openup2\jot \caja
        \ialign{\strut \hfil$\displaystyle{##}$&$
        \displaystyle{{}##}$\hfil\crcr#1\crcr}}\,}
\newif\ifdtup

\makeatletter
\def\section{\@startsection{section}{1}{\z@}
        {3ex plus-1ex minus-.2ex}{1pt plus1pt}{\large\sf\bfseries\boldmath}}
\def\subsection{\@startsection{subsection}{2}{\z@}
         {1.5ex plus-1ex minus-.2ex}{0.01pt plus1pt}{\sf\slshape}}
\def\subsubsection{\@startsection{subsubsection}{3}{\z@}
          {1.5ex plus-1ex minus-.2ex}{0.01pt plus0.2pt}{\sf\boldmath}}
\def\paragraph{\@startsection{paragraph}{4}{\z@}
           {.75ex \@plus.5ex \@minus.2ex}{-2mm}{\sf\bfseries\boldmath}}
\makeatother

 \allowdisplaybreaks
 \seceq

\usepackage{lipsum}
\usepackage{listings}
\definecolor{MyDarkGreen}{rgb}{0.0,0.4,0.0} 
\begin{document}

\thispagestyle{empty}
\noindent{\small
\hfill{  \\ 
$~~~~~~~~~~~~~~~~~~~~~~~~~~~~~~~~~~~~~~~~~~~~~~~~~~~~~~~~~~~~~~~~~$
$~~~~~~~~~~~~~~~~~~~~~~~~~~~~~~~~~~~~~~~~~~~~~~~~~~~~~~~~~~~~~~~~~$
{}
}}
\vspace*{8mm}
\begin{center}
{\large \bf
Properties of HYMNs in Examples of    \\[2pt]
Four-Color, Five-Color, and  \\[2pt]
Six-Color Adinkras   \\[2pt]
}   \vskip1in
{\large {
S.\ James Gates, Jr.\footnote{sylvester${}_-$gates@brown.edu}$^{a}$},
Yangrui Hu\footnote{yangrui\_hu@brown.edu}$^{a}$, and Kory Stiffler\footnote{kory\_stiffler@brown.edu}$^{a,b}$
}
\\*[12mm]
\emph{
\centering
{\it ${}^a$ Brown Theoretical Physics Center and Department of Physics}\\
{\it Brown University, Providence, RI 02912-1843, USA }
\\[0.2pt]
\vspace*{8mm}
{\it ${}^b$ Department of Physics and Astronomy}\\
{\it   The University of Iowa,   Iowa City, IA 52242, USA}
}
 \\*[40mm]
{ ABSTRACT}\\[4mm]
\parbox{142mm}{\parindent=2pc\indent\baselineskip=14pt plus1pt
The mathematical concept of a ``Banchoff index'' associated with
discrete Morse functions for oriented triangular meshes has been
shown to correspond to the height assignments of nodes in adinkras.
In recent work there has been introduced the concept of ``Banchoff 
matrices'' leading to HYMNs - height yielding matrix numbers. 
HYMNs map the shape of an adinkra to a set of eigenvalues 
derived from Banchoff matrices.  In the context of some examples
of four-color, minimal five-color, and minimal six-color adinkras, properties
of the HYMNs are explored. }
 \end{center}
\vfill
\noindent PACS: 11.30.Pb, 12.60.Jv\\
Keywords: adinkra, supersymmetry
\vfill
\clearpage
%

%
\section{Introduction}
\label{sec:NTR0}

Graphs have demonstrated an unexpected power to ``clear out the mathematical
underbrush" encountered by theoretical physicists.  Feynman Diagrams are
a spectacular example of this.  While it is difficult to recall, there was a time when 
Feynman Diagrams were not held in high regard.  This all changed rapidly when 
a sufficient number of physicists employed their power to efficiently calculate 
quantum corrections to physical processes. 

While we make no similar claims about breadth of possible impacts from the developments 
which began with the recognition of the existence of the ${\cal GR}$(d, $\cal N$) 
(``Garden Algebras")   \cite{GRana1,GRana2}) as a foundation of supersymmetric
representation theory and their evolution into the introduction of adinkras \cite{Adnk1},
we do hold that adinkras provide similarly important tools within the domain of the 
representation theory of supersymmetrical theories.  One hint about this involves 
the pathways \cite{codes1,codes2,codes3} that adinkras have opened from 
supersymmetrical theories, including field theories, to error-correction codes.
 
There is a substantial and rapidly growing literature on the relation of quantum
error-correction codes \cite{SHR} to the very structure of space-time (e.\ g.\ \cite{ADH})
itself.  Over and above this particular focus, there is the related similar discussion
underway regarding quantum entanglement and space-time.  Literally, it is bona 
fide query to ask, ``Is space-time a quantum error-correction code?''  However, from 
the perspective of field theory, the fields themselves are primary 
dynamical entities.  This suggests the similar question, ``Are there relations between 
fields that describe physical reality and quantum error-correction codes?''

To our knowledge, the works in \cite{codes1,codes2,codes3} are the singular
ones that hint at such a linkage.  To be clear, there have been {\em {no}} {\em {claims}} 
that these works provide for a role for {\em {quantum}} error-correction codes.
The implication of the works in \cite{codes1,codes2,codes3} is {\em {all}} {\em {
irreducible}} {\em {supersymmetric}}  {\em {field}}  {\em {theory}}  {\em {representations}}
{\em {in}}  {\em {four}}  {\em {or}}  {\em {dimensions}}  {\em {involve}}  {\em {classical}}    
 {\em {error}}-{\em {corrections}}  {\em {codes}}.   
However, it also true that the $\bm {\rm L}$-matrices and their inverses the
$\bm {\rm R}$-matrices that arise from
adinkras are generalizations of Pauli matrices.  Looking at many discussions of
quantum error-correction codes, the Pauli matrices play an important role.

Among the most primitive of adinkras are those which have the structure of nodes 
{\em {only}} occurring at two distinct height.  These are called ``valise adinkras'' with 
all bosonic nodes at the same value of height and all fermion nodes at the same (but 
different from the bosonic one) height.  The fact that the nodes representing functions 
can be differentiated or integrated is reflected in a change of nodal height in 
adinkras \cite{ulTRA}.  In the works of \cite{EH1} - \cite{Adnk2d}, various prescriptions 
have been presented for ways to map the graphical representations into numerical 
data.  While some of these involve the use of eigenvalues for this purpose, we
recently used a modified formulation\cite{HYMN1} (called Height Yielding Matrix
Numbers - HYMNs) also using eigenvalues. The ``HYMNs" approach encode the heights
of various nodes into diagonal matrices.  This approach called, ``dressing,'' was introduced in work by Toppan et.\ al.\cite{Tp1,Tp2,Tp3,Tp4,Tp5}. 
In the discussion to follow we 
explore properties of the HYMNs through examples of four-color, five-color, and 
six-color adinkras.

Works by mathematicians \cite{adnkGEO1,adnkGEO2} have connected adinkras 
to algebraic geometry and in particular Riemann surfaces.  The concept of discrete 
Morse functions for oriented triangular meshes on Riemann surfaces was introduced
into the mathematical literature some time ago \cite{B1}.  Thus, nodal heights in
adinkras are associated with values of these Morse functions.  The latter are
piecewise-linear over the meshes constructed by linear extension across edges 
and faces of the plaquettes associated with the meshes.  The bottom line is
the height assignment of any node in an adinkra corresponds to the integer
value of the Morse function and the height assignment can be ``Banchoff index'' 
of the node.   The fact that adinkra contain many nodes leads to matrices of
these indices that we call  ``Banchoff Matrices.''  The Height Yielding
Matrix Numbers - HYMNs - are the eigenvalue of these matrices.   

HYMNs provide to an intrinsic definition of the shape of an adinkra.   In a vaguely 
reminiscent way of how the Riemann curvature tensor provides a definition of 
intrinsic curvature for hyper surfaces, HYMNs provide an intrinsic way to define 
the shape of any adinkra.  In our conclusions section, a further discussion of the 
importance of this will be covered more extensively.

\newpage
\section{HYMNs: Height Yielding Matrix Numbers}
\label{sec:DeFNs}

Here, we generalize the notion of HYMNs~\cite{HYMN1} from those with only bosonic nodes 
lifted to also include lifting of fermion nodes.  Supersymmetric transformation laws encoded by 
valise adinkras can be described succinctly by the equations
\be \eqalign{
{\rm D}{}_{{}_{\rm I}} \Phi_{i} ~=~ i \, \left( {\rm L}{}_{{}_{\rm I}} \right) 
{}_{i \, {\hat k}}  \, \, \Psi_{\hat k}  ~~~,~~~
{\rm D}{}_{{}_{\rm I}} \Psi_{\hat k} ~=~ \left( {\rm R}{}_{{}_{\rm I}} \right)
{}_{{\hat k} \, i}  \, \pa_{0} \, \Phi_{i}  ~~~,
}  \label{e:1DSUSY}
\ee  
and the $N$ ${\bm {\rm L}}{}_{{}_{\rm I}}$ and $N$ ${\bm {\rm R}}{}_{{}_{\rm I}}$ matrices satisfy 
the algebra of general, real matrices describing $N$ supersymmetries between $d$ bosons and 
$d$ fermions, the so-called \emph{Garden Algebra}:
\begin{align}
{\bm {\rm L}}{}_{{}_{\rm I}} \, {\bm {\rm R}}{}_{{}_{\rm J}}  ~+~ {\bm {\rm L}}{}_{{}_{\rm J}} \, {\bm {\rm 
R}}{}_{{}_{\rm I}} ~=&~ 2 \,  \delta_{IJ} \, {\bm {\rm I}}_d
~~~,~~~
{\bm {\rm R}}{}_{{}_{\rm I}} \, {\bm {\rm L}}{}_{{}_{\rm J}}  ~+~ {\bm {\rm R}}{}_{{}_{\rm J}} \, {\bm {\rm 
L}}{}_{{}_{\rm I}}  ~=~ 2 \,  \delta_{IJ} \, {\bm {\rm I}}_d ~~~,
\end{align}
with ${\bm {\rm I}}_d
$ the $d \times d$ identity matrix. These relations imply that off-diagonal $N$ real $2d \, \times \, 2d$
matrices ${\Hat \g}{}_{{}_{\rm I}}$ constructed from the ${\bm {\rm L}}{}_{{}_{\rm I}}$ and ${\bm {\rm R}}
{}_{{}_{\rm I}}$ matrices form a Euclidean Clifford Algebra.
    
We define the node lifting operator ${\bm M}(m,w)$ that acts on an arbitrary number of $d$ fields as
\begin{align}\label{e:pword}
{\bm M}(m,w) \equiv & \begin{pmatrix}
m^{p_1} & 0 & 0 & \dots & 0 \\
0 & m^{p_1} & 0 & \dots & 0 \\
0 & 0 & m^{p_3} & \dots & 0 \\
\vdots & \vdots & \vdots & \ddots & 0 \\
0 & 0 & 0 & \dots & m^{p_{\rd}}
\end{pmatrix} ~~~,
\\ {~} \nonumber \\
\label{e:word}
w \equiv & p_1 2^{0} + p_2 2^1 + p_3 2^{2} + \dots + p_{\rd} 2^{\rd-1}~~~,~~~\text{with}~p_i 
= 0,1
\end{align}
where the word parameter $w$ is as in~\cite{permutadnk}. We define lifted bosons $\Phi(m_B,w_B)$ and lifted fermions $\Psi(m_F,w_F)$ as
\begin{align}\label{e:LiftedBosons}
    \Phi(m_B,w_B) ~&=~ {\bm M}(m_B,w_B)\Phi\\
    \label{e:LiftedFermions}
    \Psi(\m_F,w_F) ~&=~ {\bm M}(\m_F,w_F)\Psi
\end{align}
In~\cite{HYMN1}, only lifting bosonic nodes was considered. As ${\bm M}(1,w) = {\bm M}(m,0) = {\bm {\rm I}}_d$,  lifting only bosonic nodes amounts to setting either $m_F = 1$ or $w_F=0$ in the following. 

Multiplying Eqs.~\eqref{e:1DSUSY} by node lifting matrices ${\bm M}(m_B,w_B)$ and ${\bm M}(\mu_F,w_F)$ and inserting factors of ${\bm {\rm I}}_d = {\bm M}(m^{-1}_F,w_F){\bm M}(m_F,w_F)$
and ${\bm {\rm I}}_d = {\bm M}(\mu^{-1}_B,w_B) {\bm M}(\mu_B,w_B)$ results in the transformation laws
\begin{align}
	{\rm D}{}_{{}_{\rm I}}  \Phi(m_B,w_B) = i  {\bm {\rm L}}{}_{{}_{\rm I}}(m_B,m_F,w_B,w_F) \Psi(m_F,w_F) ~~~, 
	\label{eq:ModB}
\end{align}
\begin{align}
	{\rm D}{}_{{}_{\rm I}}  \Psi(\mu_F,w_F) =  {\bm {\rm R}}{}_{{}_{\rm I}}(\mu_B,\mu_F,w_B,w_F) \, \pa_0 {\Phi}(\mu_B,w_B) ~~~, ~~~~~
\label{eq:ModF}	
\end{align}
where
\begin{align}
     {\bm {\rm L}}{}_{{}_{\rm I}}(m_B,m_F,w_B,w_F) ~&=~ {\bm M}(m_B,w_B){\bm  {\rm L}}{}_{{}_{\rm I}}{\bm M}(m_F^{-1},w_F)\label{equ:L_liftBF}\\
    {\bm  {\rm R}}{}_{{}_{\rm I}}(\mu_B,\mu_F,w_B,w_F) ~&=~ {\bm M}(\mu_F,w_F) {\bm {\rm R}}{}_{{}_{\rm I}}{\bm M}(\mu_B^{-1},w_B)\label{equ:R_liftBF}
\end{align}
The way we have written these equations, $\mu$ is not specific to fermions and $m$ is not specfic to bosons. Rather, $\mu$ is specific to the $\bm {\rm R}$-matrices and $m$ to the $\bm {\rm L}$-matrices. See Eqs. (\ref{eq:ModB}) and (\ref{eq:ModF}).

Based on equation (\ref{equ:L_liftBF}) and (\ref{equ:R_liftBF}), it's clear that the final $\bm {\rm L}$ and $\bm {\rm R}$ matrices after lifting both bosons and fermions do \textit{not} depend on the order of lifting operations. Since two diagonal matrices commute with each other and matrix multiplication is associative.
Namely, there is only one unique set of L and R matrices corresponding to a specified adinkra, no matter whether it is valise or non-valise. 

The redefined matrices ${\bm {\rm L}}{}_{{}_{\rm I}}(m,w)$ and ${\bm {\rm R}}{}_{{}_{\rm I}}(\mu,w)$ satisfy the $GR(d,N)$ algebra in the 
$\mu_F \to m_F$ and $\mu_B \to m_B$ limit:
\begin{align}
& {\bm {\rm L}}{}_{{}_{\rm I}}(m_B,m_F,w_B,w_F)  {\bm {\rm R}}{}_{{}_{\rm J}}(\mu_B,\mu_F,w_B,w_F) +   {\bm {\rm L}}{}_{{}_{\rm J}}(m_B,m_F,w_B,w_F)  {\bm {\rm R}}{}_{{}_{\rm I}}(\mu_B,\mu_F,w_B,w_F)  \cr
&= {\bm M}(m_B,w_B)\left[ {\bm {\rm L}}{}_{{}_{\rm I}} {\bm M}(\mu_F/m_F,w_F) {\bm {\rm R}}{}_{{}_{\rm J}} +  {\bm {\rm L}}{}_{{}_{\rm J}}{\bm M}(\mu_F/m_F,w_F ) {\bm {\rm R}}{}_{{}_{\rm I}} \right] {\bm M}(\mu^{-1}_B,w_B)  \cr
&\to {\bm M}(m_B,w_B)\left[ {\bm {\rm L}}{}_{{}_{\rm I}}  {\bm {\rm R}}{}_{{}_{\rm J}} +  {\bm {\rm L}}{}_{{}_{\rm J}} {\bm {\rm R}}{}_{{}_{\rm I}}  \right] {\bm M}(\mu^{-1}_B,w_B)   = \,2 \, \delta_{{\rm I} \, {\rm J}} \,  {\bm M}(m_B/\mu_B,w_B) ~~~,~~~ \text{for}~\mu_F \to m_F \cr
 &\to \,  2\,  \delta_{{\rm I} \, {\rm J}}\, {\bm {\rm I}}{}_d ~~~,~~~ \text{for}~\mu_B \to m_B
\end{align}
The same results hold in the $\mu_B \to m_B$ and $\mu_F \to m_F$ limit for the $ {\bm {\rm R}}{}_{{}_{\rm I}} {\bm {\rm L}}{}_{{}_{\rm J}} +  {\bm {\rm R}}{}_{{}_{\rm J}} {\bm {\rm L}}{}_{{}_{\rm I}}$ 
algebra
\begin{align}
	& {\rm R}{}_{{}_{\rm I}}(\mu_B,\mu_F,w_B,w_F)   {\rm L}{}_{{}_{\rm J}}(m_B,m_F,w_B,w_F)+   {\rm R}{}_{{}_{\rm J}}(\mu_B,\mu_F,w_B,w_F)   {\rm L}{}_{{}_{\rm I}}(m_B,m_F,w_B,w_F) \cr
	&\to   2\,  \delta_{{\rm I} \, {\rm J}}\, {\bm {\rm I}}{}_d ~~~,~~~ \text{for}~\mu_B \to m_B~~~,~~~\text{and}~~~~\mu_F \to m_F
\end{align}

The $m/\mu$ ratio will be prevalent in the rest of our calculations, so we define\begin{align}
 \rho_B = m_B/\mu_B~~~,~~~ \rho_F = \m_F/m_F~~~. 
\end{align}
For $N=4$, lifted Banchoff $\bm B$-matrices are defined as:
\begin{equation}
    \begin{split}
        &{\bm B}_L(\rho_B,\rho_F,w_B,w_F) ~=~\\ &{\bm  {\rm L}}{}_{{}_{4}}(m_B,m_F,w_B,w_F){\bm  {\rm R}}{}_{{}_{3}}(\mu_B,\mu_F,w_B,w_F){\bm  {\rm L}}{}_{{}_{2}}(m_B,m_F,w_B,w_F){\bm  {\rm R}}{}_{{}_{1}}(\mu_B,\mu_F,w_B,w_F)\\
    &{\bm B}_R(\rho_B,\rho_F,w_B,w_F) ~=~\\ &{\bm  {\rm R}}{}_{{}_{4}}(\mu_B,\mu_F,w_B,w_F){\bm  {\rm L}}{}_{{}_{3}}(m_B,m_F,w_B,w_F){\bm  {\rm R}}{}_{{}_{2}}(\mu_B,\mu_F,w_B,w_F){\bm  {\rm L}}{}_{{}_{1}}(m_B,m_F,w_B,w_F)
    \label{eq:BLBRdef}
    \end{split}
\end{equation}

We will define Banchoff matrices for $N=5$ and $N=6$ in the following sections.

In the remainder of this work, we will demonstrate, in the context of specific example,
some properties of these constructions.  Our examples are chosen from systems of four color, five
color, and then six color adinkras.

\newpage

\section{GR(4,4) Calculations}

Since adinkras can be obtained from the dimensional reduction of supersymmetric
theories from higher dimensions, there are multiple higher dimensional starting
points.  Thus, models with the same number of independent supercharges under
such reduction lead to the same adinkras.  In particular, 4D, $\cal N$ = 1 \cite{Gates:2009me}, 2D, 
$\cal N$ = (2,2) \cite{GHR}, 2D, $\cal N$ = (4,0) \cite{DGR} and 1D, $\cal N$ = 4 supersymmetrical theories 
\cite{G-1}
must produce adinkras that lie in the set of the 36,864 adinkras associated
with the Coxeter Group $BC_4$.

If the starting point is chosen in the domain of 4D, $\cal N$ = 1 supersymmetrical 
theories, among the most familiar models involve the chiral, vector and tensor 
supermultiplets.  The explicit forms of the $\bm {\rm L}$-matrices and $\bm {\rm R}$-matrices
for the 4D, $\cal N$ = 1 supermultiplets can be found in the work seen in \cite{Gates:2009me}. 
This is our starting point discussed below. In the following sections, we not only present HYMNs for various supermultiplets, but also HYMNs if we drop all dashings in the associated adinkra.  

\subsection{Chiral Supermultiplet:}

\begin{figure}[htp!]
    \centering
    \includegraphics[width=0.4\textwidth]{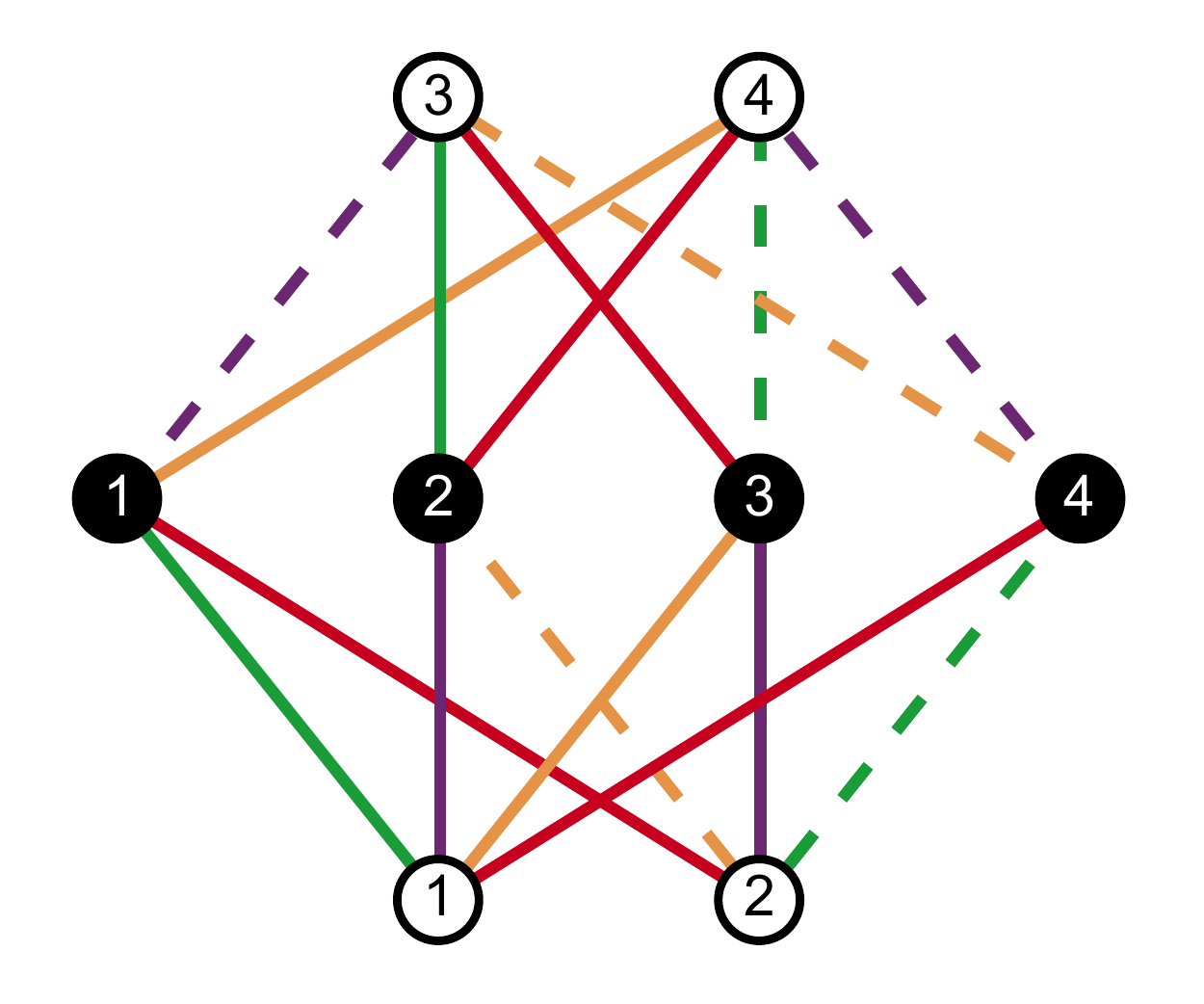}
    \caption{Chiral Supermultiplet Adinkra}
    \label{fig:CMadink}
\end{figure}

In Fig.\ \ref{fig:CMadink}, there is shown the adinkra for chiral supermultiplet.

\begin{table}[htp!]
\centering
\begin{tabular}{|c|c|}
\hline
${\bm B}_L$ eigenvalues   &   ${\bm B}_R$ eigenvalues  \\\hline\hline
$\{  \, \rho_B \, ,  \rho_B \, ,  \rho_B \, ,  \rho_B \, \}$ & 
$\{ \, - \, 1 \, ,  - 1 \, , - \, \rho{}^{2}_B \, ,  - \, \rho{}^{2}_B \, \}$  \\\hline
${\bm B}_L$ eigenvalues (dropping all dashings)   &   
${\bm B}_R$ eigenvalues (dropping all dashings) \\\hline\hline 
$\{  \, \rho_B \, ,  \rho_B \, ,  \rho_B \, ,  \rho_B \, \}$ & 
$\{ \, 1 \, ,   1 \, ,  \rho{}^{2}_B \, ,  \rho{}^{2}_B \, \}$  \\\hline 
\end{tabular}
\caption{Results of ${\bm B}_L$ and ${\bm B}_R$ eigenvalues with/without 
dashings for the adinkra in Fig \ref{fig:CMadink}}
\label{tab:result_GR(4,4)CM}
\end{table}

\subsection{Vector Supermultiplet:}

\begin{figure}[htp!]
    \centering
    \includegraphics[width=0.4\textwidth]{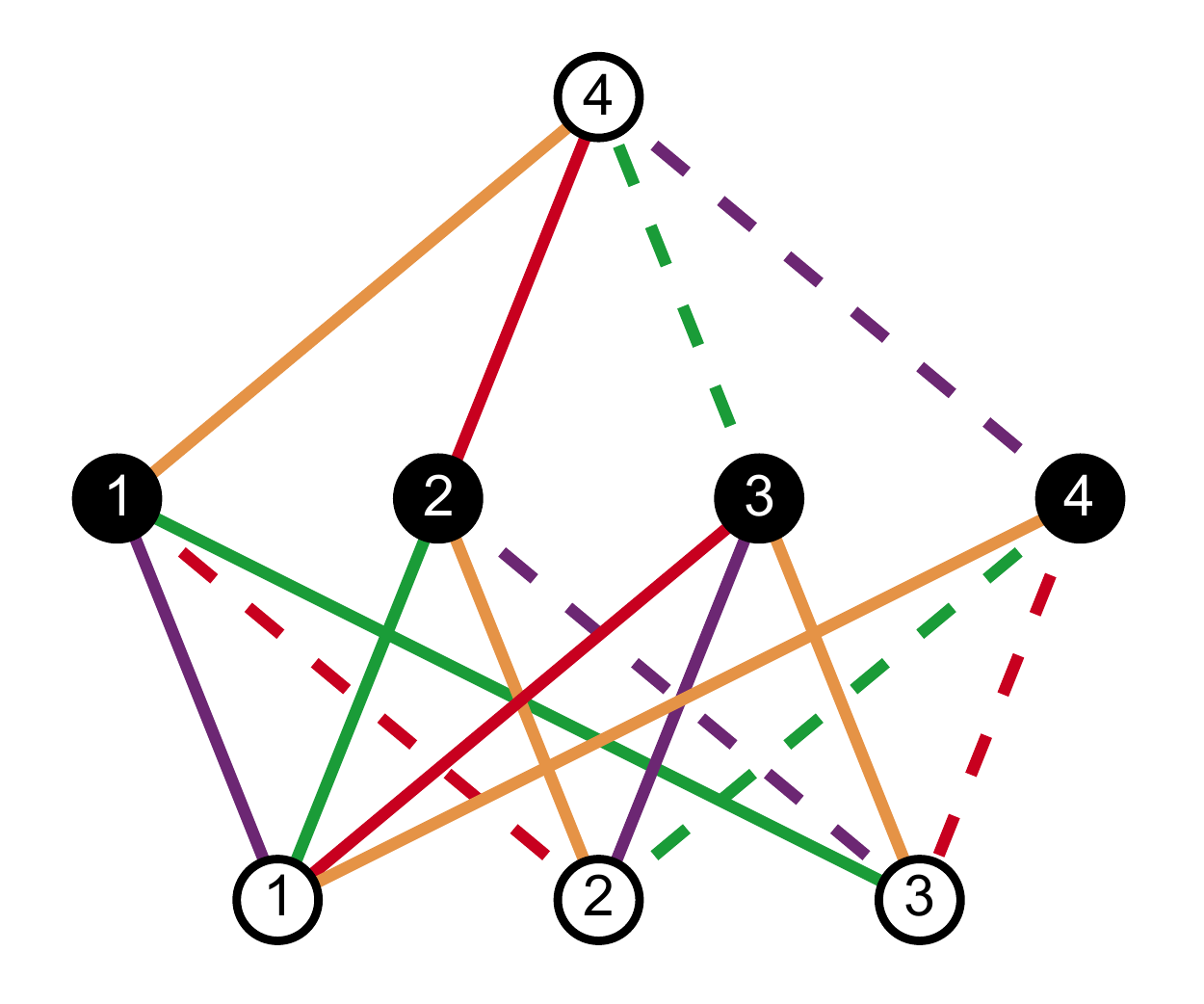}
    \caption{Vector Supermultiplet Adinkra}
    \label{fig:VMadink}
\end{figure}

In Fig.\ \ref{fig:VMadink}, there is shown the adinkra for vector supermultiplet.

\begin{table}[htp!]
\centering
\begin{tabular}{|c|c|}
\hline
${\bm B}_L$ eigenvalues   &   ${\bm B}_R$ eigenvalues  \\\hline\hline
$\{  \, - \, 1 \, , -\, 1 \, , -\, \rho_B \, , -\, \rho_B  \, \}$ & 
$\{ \, 1 \, , 1,\, \rho_B \, , \rho_B \, \}$  \\\hline
${\bm B}_L$ eigenvalues (dropping all dashings)   &   
${\bm B}_R$ eigenvalues (dropping all dashings) \\\hline\hline 
$\{  \, 1 \, , 1 \, , \rho_B \, , \rho_B \, \}$ & 
$\{ \, 1 \, , 1,\, \rho_B \, , \rho_B \, \}$   \\\hline 
\end{tabular}
\caption{Results of ${\bm B}_L$ and ${\bm B}_R$ eigenvalues with/without 
dashings for the adinkra in Fig \ref{fig:VMadink}}
\label{tab:result_GR(4,4)VM}
\end{table}

\newpage

\subsection{Tensor Supermultiplet:}

\begin{figure}[htp!]
    \centering
    \includegraphics[width=0.4\textwidth]{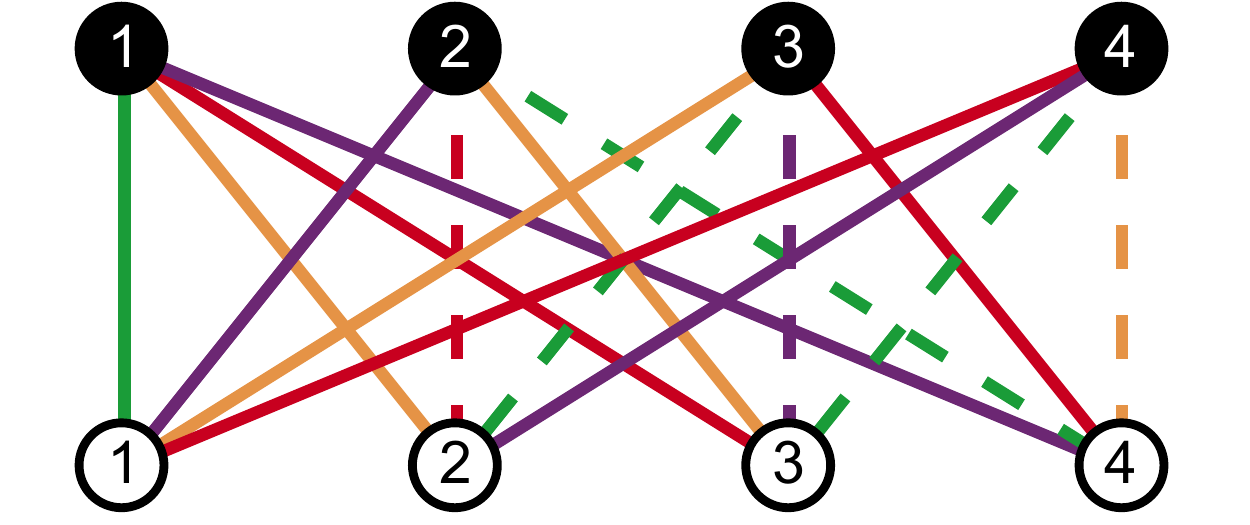}
    \caption{Tensor Supermultiplet Adinkra}
    \label{fig:TMadink}
\end{figure}

In Fig.\ \ref{fig:TMadink}, there is shown the adinkra for tensor supermultiplet.

\begin{table}[htp!]
\centering
\begin{tabular}{|c|c|}
\hline
${\bm B}_L$ eigenvalues   &   ${\bm B}_R$ eigenvalues  \\\hline\hline
$\{  \,  - \, 1 \, ,  - 1 \, , - \, 1 \, ,   - \, 1 \, \}$ & 
$\{ \,  1 \, ,  1,  \,   1,  1  \, \}$  \\\hline
${\bm B}_L$ eigenvalues (dropping all dashings)   &   
${\bm B}_R$ eigenvalues (dropping all dashings) \\\hline\hline 
$\{ \,  1 \, ,  1,  \,   1,  1  \,  \}$ & 
$\{ \,  1 \, ,  1,  \,   1,  1 \, \}$  \\\hline 
\end{tabular}
\caption{Results of ${\bm B}_L$ and ${\bm B}_R$ eigenvalues with/without 
dashings for the adinkra in Fig \ref{fig:TMadink}}
\label{tab:result_GR(4,4)TM}
\end{table}

\subsection{Four Supermultiples: SM-I, SM-II, SM-III, SM-IV}

\begin{figure}[htp!]
\centering
\begin{minipage}{0.46\textwidth}
    \centering
    \includegraphics[width=0.9\textwidth]{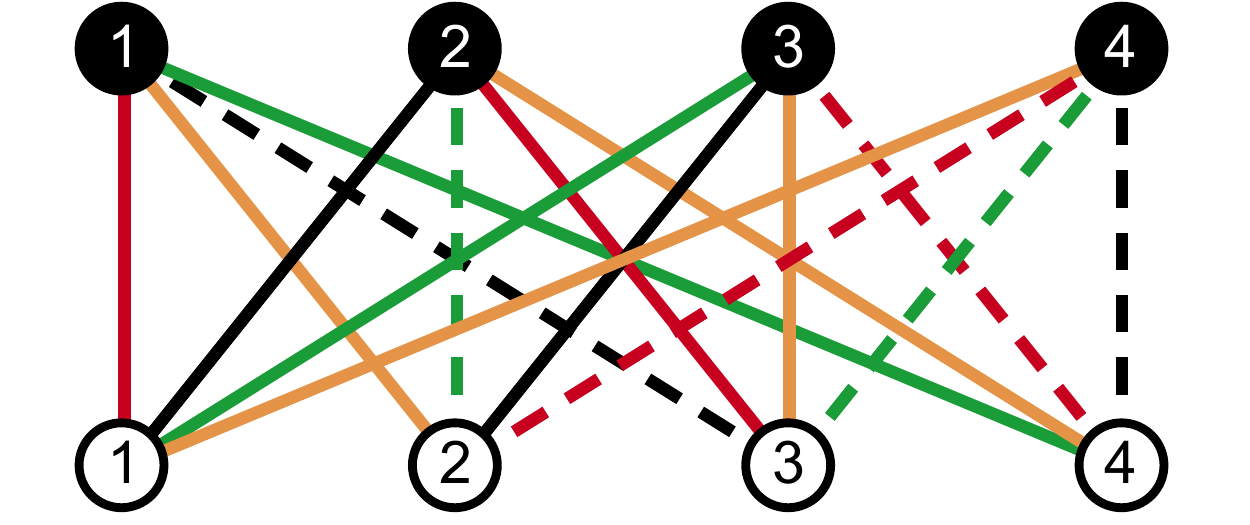}
    \caption{Adinkra Diagram for SM-I}
    \label{Fig:SM-I}
\end{minipage}
\begin{minipage}{0.46\textwidth}
    \centering
    \includegraphics[width=0.9\textwidth]{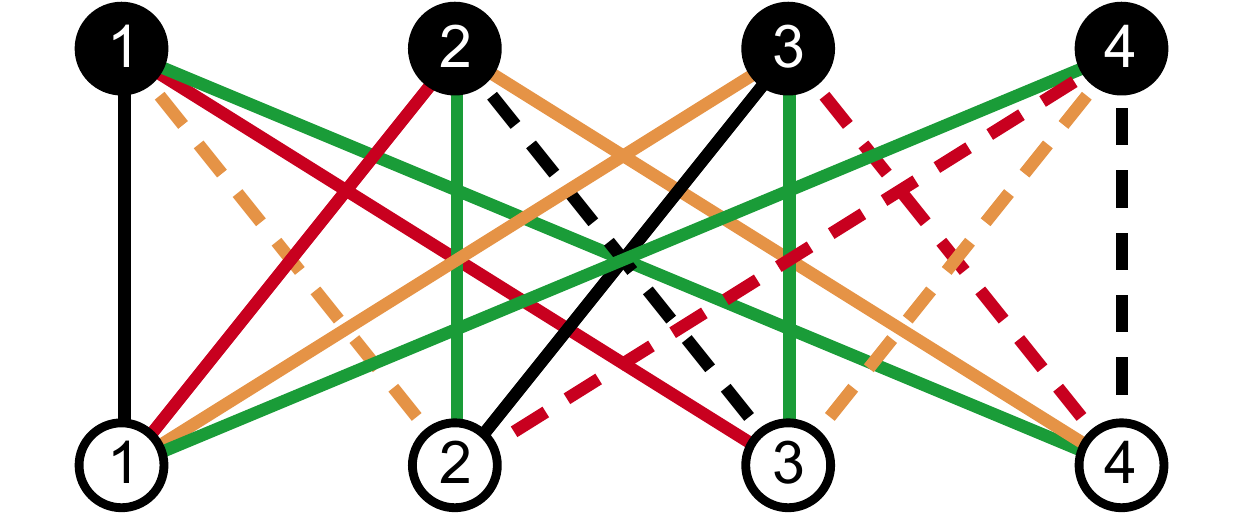}
    \caption{Adinkra Diagram for SM-II}
   \label{Fig:SM-II}
\end{minipage}
\end{figure}

\begin{figure}[htp!]
\centering
\begin{minipage}{0.46\textwidth}
    \centering
    \includegraphics[width=0.9\textwidth]{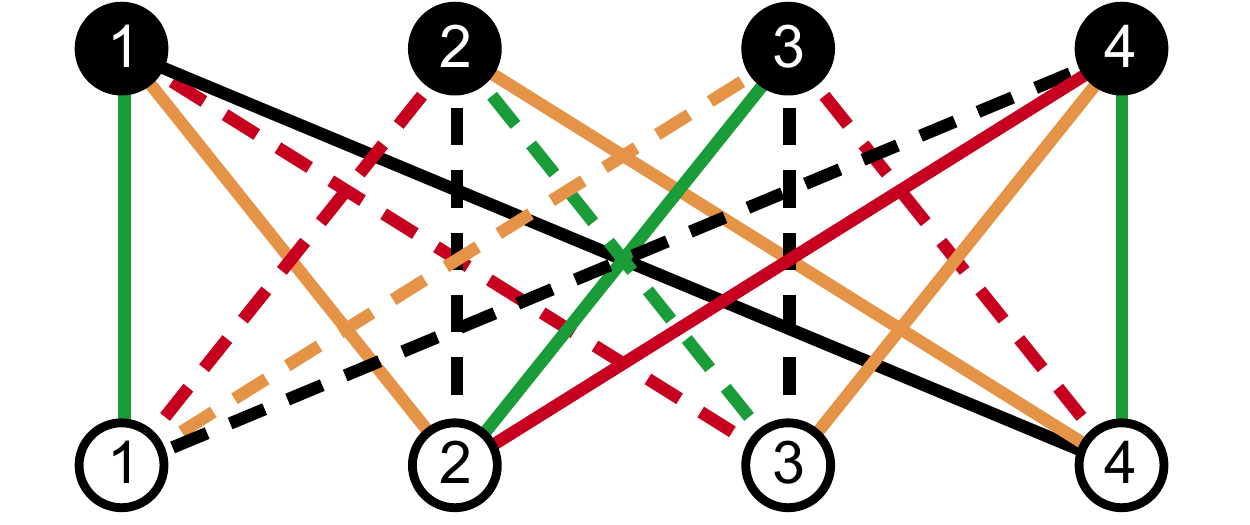}
    \caption{Adinkra Diagram for SM-III}
    \label{Fig:SM-III}
\end{minipage}
\begin{minipage}{0.46\textwidth}
    \centering
    \includegraphics[width=0.9\textwidth]{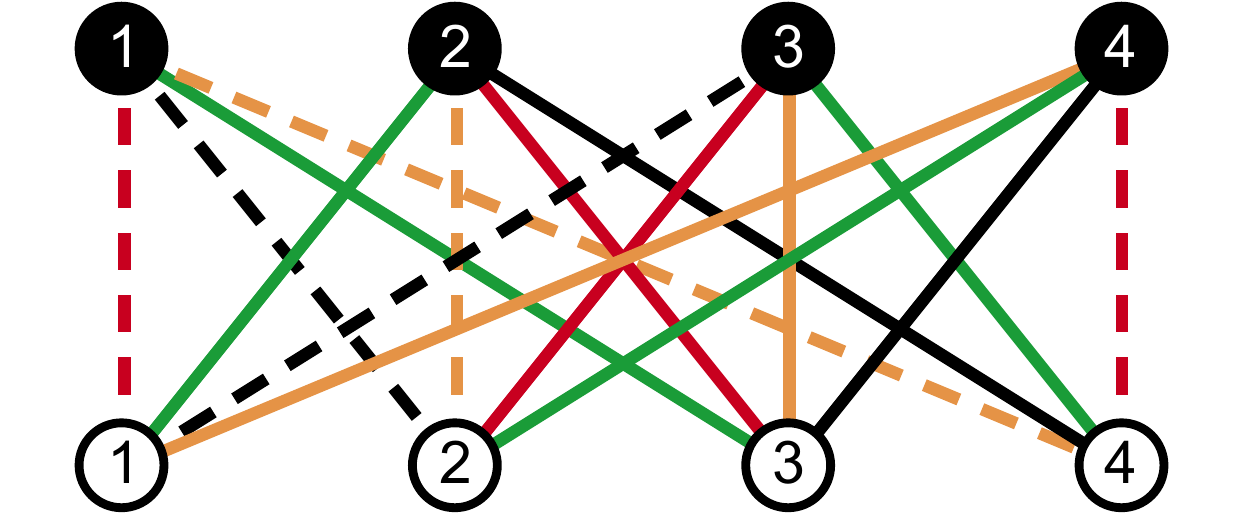}
    \caption{Adinkra Diagram for SM-IV}
   \label{Fig:SM-IV}
\end{minipage}
\end{figure}

In this subsection, the valise adinkras associated with the domain of 
2D, $\cal N$ = (4,0) supersymmetric models are considered below.
The explicit forms of the $\bm {\rm L}$-matrices and $\bm {\rm R}$-matrices
for the 2D, $\cal N$ = (4,0) and 1D, $\cal N$ = 4 supermultiplets can be found 
in the work seen in \cite{G-1}.
Thus in Fig.\ \ref{Fig:SM-I} through \ref{Fig:SM-IV}, there are shown 
the adinkras for supermultiplets SM-I to SM-IV.

These four adinkras, as well as all 36,864 valise GR(4,4) adinkras share the same eigenvalues. 
Eigenvalues for ${\bm B}_L$ matrix are:
\begin{table}[htp!]
\centering
\begin{tabular}{|c|c|}
\hline
${\bm B}_L$ eigenvalues   &   ${\bm B}_R$ eigenvalues  \\\hline\hline
$\{  \, \chi{}_{\rm o}, \,  \chi{}_{\rm o}, \,  \chi{}_{\rm o}, \,  \chi{}_{\rm o}  \, \}$ & 
$\{ \, - \, \chi{}_{\rm o},  \, - \, \chi{}_{\rm o},  \,   - \, \chi{}_{\rm o},  \,   - \, \chi{}_{\rm o}  
\, \}$  \\\hline
${\bm B}_L$ eigenvalues (dropping all dashings)   &   
${\bm B}_R$ eigenvalues (dropping all dashings) \\\hline\hline 
$\{  \, 1, \,    1, \,   1, \,    1 \, \}$ & 
$\{ \, 1, \,    1, \,   1, \,    1  \, \}$  \\\hline 
\end{tabular}
\caption{Results of ${\bm B}_L$ and ${\bm B}_R$ eigenvalues with/without 
dashings for the 36, 864 valise adinkras \newline $~~~~~~~~~~~~~\,$
 associated with (4, 0) SUSY}
\label{tab:result_GR(4,4)x}
\end{table}
where 
\begin{equation}
    \chi_o({\mathcal R}) ~=~ \begin{cases}
    1 ~~~~~\text{when }\mathcal{R}=\text{SM-I, SM-IV}\\
    -1 ~~~\text{when }\mathcal{R}=\text{SM-II, SM-III}\\
    \end{cases}
\end{equation}
which is true for all the valise GR(4,4) adinkras.

\subsection{The Collapse of the Variant Representations}
\label{sec:2-Vars}

There are ten off-shell 4D, $\cal N$ = 1 off-shell supermultiplets.  For completeness (as well
as correcting some sign errors in previous publications) the complete description of the component 
fields as well as SUSY transformation laws are given in Appendix \ref{appen:transf_law}.  

The point is to note that any theory that has a description in terms of 4D, $\cal N$ = 1 off-shell 
supermultiplet can be reduced to a 1D, $N$ = 4 off-shell supermultiplet which may be transformed
into a valise adinkra by bring all the bosonic fields to the same level in the corresponding adinkra.

We begin by briefly giving
the names and component fields listed in each of these supermultiplets.  We have for all those 
supermultiplets related to the chiral supermultiplet the following:

\noindent
(1.) $ {\rm {Chiral~Supermultiplet}: ~(A, \, B, \,  \psi_a , \, F, \, G)}$ 
\newline \noindent
(2.) $ {\rm {Hodge-Dual~ \#1~Chiral~Supermultiplet}: ~(A, \, B, \,  \psi_a , \, {\rm f}_{\mu
 \, \nu \, \rho}, \, G)}$
\newline \noindent
(3.) $ {\rm  {Hodge-Dual~ \#2~Chiral~Supermultiplet:} ~(A, \, B, \,  \psi_a , \, F, \, {\rm 
g}_{\mu \, \nu \, \rho})}$
\newline \noindent
(4.) $ {\rm  {Hodge-Dual~ \#3~Chiral~Supermultiplet:} ~(A, \, B, \,  \psi_a , \, 
{\rm f}_{\mu \, \nu \, \rho}, \, {\rm g}_{\mu \, \nu \, \rho})}$

\begin{table}[htp!]
\centering
\begin{tabular}{|c|c|}
\hline
${\bm B}_L$ eigenvalues   &   ${\bm B}_R$ eigenvalues  \\\hline\hline
$\{  \, 1 \, , 1 \, ,  \rho_B \, ,  \rho_B \, \}$ & 
$\{ \, - \, 1 \, ,  - 1 \, , - \, \rho_B \, ,  - \, \rho_B \, \}$  \\\hline
${\bm B}_L$ eigenvalues (dropping all dashings)   &   
${\bm B}_R$ eigenvalues (dropping all dashings) \\\hline\hline 
$\{  \, 1 \, , 1 \, ,  \rho_B \, ,  \rho_B \, \}$ & 
$\{  \, 1 \, , 1 \, ,  \rho_B \, ,  \rho_B \, \}$  \\\hline 
\end{tabular}
\caption{Results of ${\bm B}_L$ and ${\bm B}_R$ eigenvalues with/without 
dashings for Hodge-Dual \#1 and Hodge- \newline $~~~~~~~~~~~~~$
Dual \#2 Chiral Supermultiplet}
\label{tab:result_HodgeDual-1CM}
\end{table}

The eigenvalues shown in tables \# \ref{tab:result_HodgeDual-1CM} and \# \ref{tab:result_HodgeDual-3CM} result with the four dimensional supermultiplet
is case into the form of a 1D valise supermultiplet.

\begin{table}[htp!]
\centering
\begin{tabular}{|c|c|}
\hline
${\bm B}_L$ eigenvalues   &   ${\bm B}_R$ eigenvalues  \\\hline\hline
$\{  \, 1 \, , 1 \, ,  1 \, ,  1 \, \}$ & 
$\{ \, -1 \, ,  - 1 \, , - 1 \, ,  -1 \, \}$  \\\hline
${\bm B}_L$ eigenvalues (dropping all dashings)   &   
${\bm B}_R$ eigenvalues (dropping all dashings) \\\hline\hline 
$\{  \, 1 \, , 1 \, ,  1 \, , 1 \, \}$ & 
$\{  \, 1 \, , 1 \, ,  1 \, , 1 \, \}$  \\\hline 
\end{tabular}
\caption{Results of ${\bm B}_L$ and ${\bm B}_R$ eigenvalues with/without 
dashings for Hodge-Dual \#3 Chiral \newline $~~~~~~~~~~~~~$
Supermultiplet}
\label{tab:result_HodgeDual-3CM}
\end{table}
It should be noted that performing a parity exchange on the component bosonic fields of the original chiral supermultiplet returns the same field content.
The reason for this is that the bosonic fields in the chiral-like supermultiplets all come in opposite parity pairs.  So performing a parity transformation on the bosons only ``swaps'' members of the parings.
This is not so for the remaining supermultiplets.

We have for all those supermultiplets related to the vector supermultiplet the following

\noindent
(1.) $ {\rm  {Vector~Supermultiplet:}~ (A{}_{\mu} , \, \l_b , \,  {\rm d})}$
\newline \noindent
(2.) $ {\rm  {Axial-Vector~Supermultiplet:}~ (U{}_{\mu} , \, {\Tilde \l}_b , \,  {\Tilde {\rm d}}
)}$
\newline \noindent
(3.) $ {\rm {Hodge-Dual~Vector~Supermultiplet:}~ (A{}_{\mu} , \, \l_b , \,  {\rm d}
{}_{\mu \, \nu \, \rho} 
)}$
\newline \noindent
(4.) $ {\rm {Hodge-Dual~ Axial-Vector~Supermultiplet:}~ ({\Tilde U}{}_{\mu} , \, {\Tilde \l}
{}_b , \,  {\Tilde {\rm d}}{}_{\mu \, \nu \, \rho} )}$

\begin{table}[htp!]
\centering
\begin{tabular}{|c|c|}
\hline
${\bm B}_L$ eigenvalues   &   ${\bm B}_R$ eigenvalues  \\\hline\hline
$\{ \, - \, 1 \, ,  - 1 \, , - \, \rho_B \, ,  - \, \rho_B \, \}$ & 
$\{  \, 1 \, , 1 \, ,  \rho_B \, ,  \rho_B \, \}$  \\\hline
${\bm B}_L$ eigenvalues (dropping all dashings)   &   
${\bm B}_R$ eigenvalues (dropping all dashings) \\\hline\hline 
$\{  \, 1 \, , 1 \, ,  \rho_B \, ,  \rho_B \, \}$ & 
$\{  \, 1 \, , 1 \, ,  \rho_B \, ,  \rho_B \, \}$  \\\hline 
\end{tabular}
\caption{Results of ${\bm B}_L$ and ${\bm B}_R$ eigenvalues with/without 
dashings for Axial-Vector Supermultiplet}
\label{tab:result_Axial-VM}
\end{table}

\begin{table}[htp!]
\centering
\begin{tabular}{|c|c|}
\hline
${\bm B}_L$ eigenvalues   &   ${\bm B}_R$ eigenvalues  \\\hline\hline
$\{  \, -1 \, , -1 \, ,  -1 \, ,  -1 \, \}$ & 
$\{ \, 1 \, ,  1 \, ,  1 \, ,  1 \, \}$  \\\hline
${\bm B}_L$ eigenvalues (dropping all dashings)   &   
${\bm B}_R$ eigenvalues (dropping all dashings) \\\hline\hline 
$\{  \, 1 \, , 1 \, ,  1 \, , 1 \, \}$ & 
$\{  \, 1 \, , 1 \, ,  1 \, , 1 \, \}$  \\\hline 
\end{tabular}
\caption{Results of ${\bm B}_L$ and ${\bm B}_R$ eigenvalues with/without 
dashings for Hodge-Dual Vector Super- \newline $~~~~~~~~~~~\,~~$
multiplet, Hodge-Dual Axial-Vector Supermultiplet, and Axial-Tensor Supermultiplet}
\label{tab:result_HodgeDual-VM}
\end{table}

Finally, we come to the tensor supermultiplet and its parity exhanged version.  Since
this supermultiplet does not possess any auxiliary fields, there is not a possibility
to perform a Hodge-type duality.  So the only possible transformation is to perform
a parity exchange on the bosons.  This leads to:

\noindent
(1.) $ {\rm  {Tensor~Supermultiplet:} ~(\varphi, \, B{}_{\mu \, \nu }, \,  \chi_a )}$
\newline \noindent
(2.) $ {\rm {Axial-Tensor~Supermultiplet:} ~( {\Tilde {\varphi}}, \, 
C{}_{\mu \, \nu }, \,  {\Tilde {\chi}} {}_a )}$

These studies suggest that two simple functions of the eigenvalues that are important for the introduction of a class-based structure for these diagrams would be the trace and determinant.
The trace and determinant have the properties that they are invariant under any permutation of the eigenvalues.
Actually, since there are two sets of eigenvalues (one set for ${\bm B}_L$ and one for ${\bm B}_R$). we are able to analyze the trace and determinant for each. These results are shown in appendix~\ref{a:EigenSumProd}.

\newpage
\section{An Example of 4-color Non-minimal Adinkra\label{sec:4colornonmininal}}

\begin{figure}[htp!]
    \centering
    \includegraphics[width=0.4\textwidth]{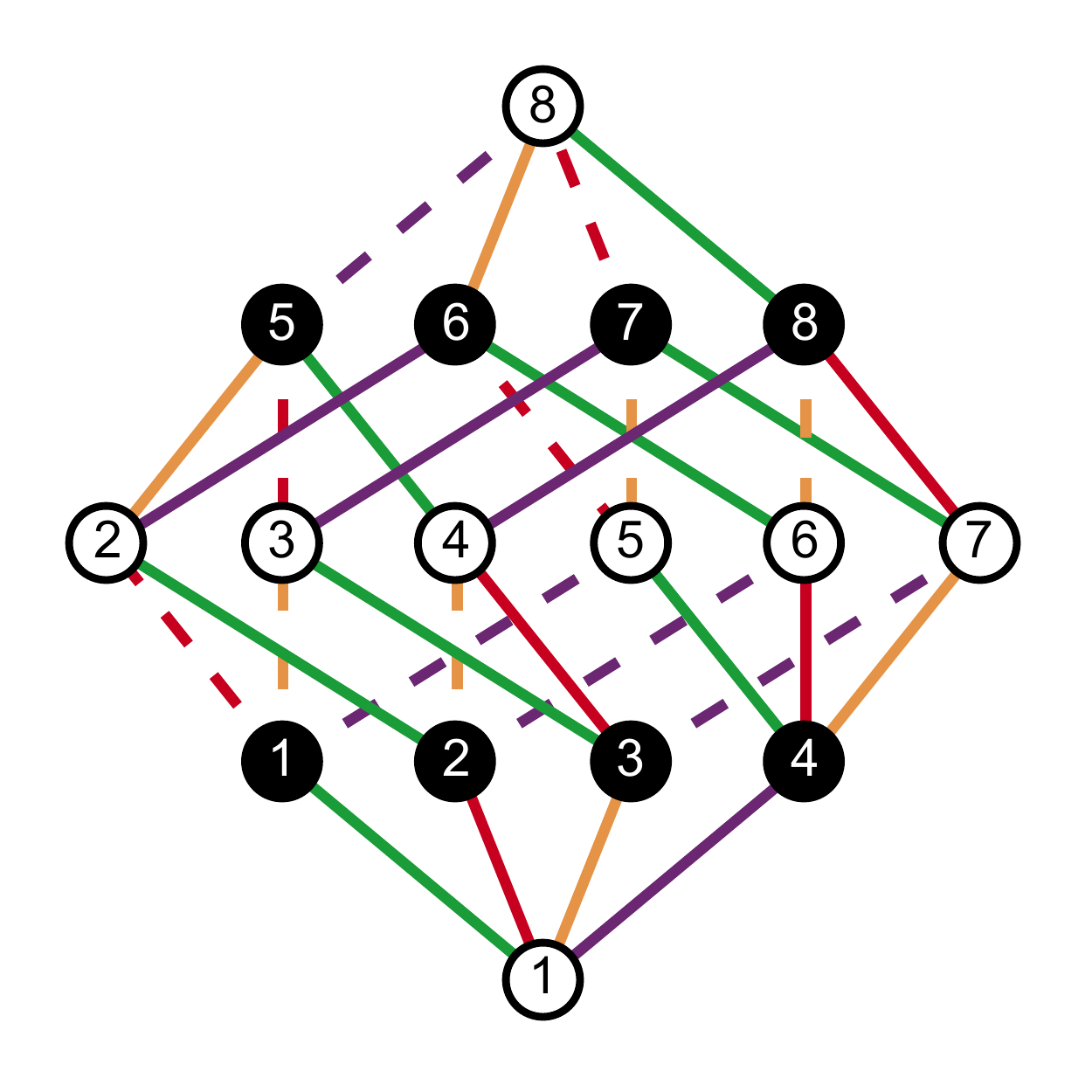}
    \caption{An Example of 4-color Non-minimal Adinkra}
    \label{fig:adink48}
\end{figure}

In this chapter, the discussion will turn to an exploration of the response of the HYMNs to changes of shape of 4-color adinkras which possess 8-closed and 8-open nodes as seen in Fig.\ \ref{fig:adink48}.  These correspond to reducible superfields. The 4D, $\mathcal{N}=1$ real scalar superfield was shown in~\cite{Gates:2011aa} to be depicted by the adinkra in Fig.~\ref{fig:adink48} similar to how the 4D, $\mathcal{N}=1$ chiral, vector, and tensor multiplets were shown in~\cite{Gates:2009me} to be depicted by the adinkras in Figs.~\ref{fig:CMadink}, \ref{fig:VMadink}, and \ref{fig:TMadink}.

Recall that the definition of Banchoff $\bm B$-matrices is:
\begin{align}
    {\bm B}_L ~=&~ {\bm  {\rm L}}{}_{{}_{4}}{\bm  {\rm R}}{}_{{}_{3}}{\bm  {\rm L}}{}_{{}_{2}}{\bm  {\rm R}}{}_{{}_{1}}~~~,\\
    {\bm B}_R ~=&~ {\bm  {\rm R}}{}_{{}_{4}}{\bm  {\rm L}}{}_{{}_{3}}{\bm  {\rm R}}{}_{{}_{2}}{\bm  {\rm L}}{}_{{}_{1}}~~~.
\end{align}

For the adinkra in Figure \ref{fig:adink48}, the corresponding ${\bm {\rm L}}$-matrices are:
\begin{align}
    {\bm  {\rm L}}{}_{{}_{1}} ~&=~ \left(
\begin{matrix}
1 & 0 & 0 & 0 & 0 & 0 & 0 & 0\\
0 & 1 & 0 & 0 & 0 & 0 & 0 & 0\\
0 & 0 & 1 & 0 & 0 & 0 & 0 & 0\\
0 & 0 & 0 & 0 & 1 & 0 & 0 & 0\\
0 & 0 & 0 & 1 & 0 & 0 & 0 & 0\\
0 & 0 & 0 & 0 & 0 & 1 & 0 & 0\\
0 & 0 & 0 & 0 & 0 & 0 & 1 & 0\\
0 & 0 & 0 & 0 & 0 & 0 & 0 & 1\\
\end{matrix}
\right) ~~~~~,~~~~~~
    {\bm  {\rm L}}{}_{{}_{2}} ~=~ \left(
\begin{matrix}
0 & 0 & 0 & 1 & 0 & 0 & 0 & 0\\
0 & 0 & 0 & 0 & 0 & 1 & 0 & 0\\
0 & 0 & 0 & 0 & 0 & 0 & 1 & 0\\
0 & 0 & 0 & 0 & 0 & 0 & 0 & 1\\
-1 & 0 & 0 & 0 & 0 & 0 & 0 & 0\\
0 & -1 & 0 & 0 & 0 & 0 & 0 & 0\\
0 & 0 & -1 & 0 & 0 & 0 & 0 & 0\\
0 & 0 & 0 & 0 & -1 & 0 & 0 & 0\\
\end{matrix}
\right) ~~,\\ 
    {\bm  {\rm L}}{}_{{}_{3}} ~&=~ \left(
\begin{matrix}
0 & 0 & 1 & 0 & 0 & 0 & 0 & 0\\
0 & 0 & 0 & 0 & 1 & 0 & 0 & 0\\
-1 & 0 & 0 & 0 & 0 & 0 & 0 & 0\\
0 & -1 & 0 & 0 & 0 & 0 & 0 & 0\\
0 & 0 & 0 & 0 & 0 & 0 & -1 & 0\\
0 & 0 & 0 & 0 & 0 & 0 & 0 & -1\\
0 & 0 & 0 & 1 & 0 & 0 & 0 & 0\\
0 & 0 & 0 & 0 & 0 & 1 & 0 & 0\\
\end{matrix}
\right) ~~,
    {\bm  {\rm L}}{}_{{}_{4}} ~=~ \left(
\begin{matrix}
0 & 1 & 0 & 0 & 0 & 0 & 0 & 0\\
-1 & 0 & 0 & 0 & 0 & 0 & 0 & 0\\
0 & 0 & 0 & 0 & -1 & 0 & 0 & 0\\
0 & 0 & 1 & 0 & 0 & 0 & 0 & 0\\
0 & 0 & 0 & 0 & 0 & -1 & 0 & 0\\
0 & 0 & 0 & 1 & 0 & 0 & 0 & 0\\
0 & 0 & 0 & 0 & 0 & 0 & 0 & 1\\
0 & 0 & 0 & 0 & 0 & 0 & -1 & 0\\
\end{matrix}
\right) ~~. 
\end{align}

In this case, we not only lift bosons but also lift fermions. We define the boson lifting matrix $M(m_B,w_B)$ and fermion lifting matrix $M(m_F,w_F)$ as in Eqs.~(\ref{e:LiftedBosons}) and~\eqref{e:LiftedFermions}, respectively. By analyzing the HYMNs, we get results listed in Table \ref{tab:result_GR(8,4)}. 
Eigenvalues are not sensitive to dashings. 

\begin{table}[htp!]
    \centering
    \begin{tabular}{|c|c|}
    \hline
    ${\bm B}_L$ eigenvalues   &   ${\bm B}_R$ eigenvalues  \\\hline\hline
       $\{  -\rho_B^2\rho_F, -\rho_B^2\rho_F, -\rho_B^2\rho_F, -\rho_B^2\rho_F, $ &  $\{  -\rho_B^2\rho_F, -\rho_B^2\rho_F, -\rho_B^2\rho_F, -\rho_B^2\rho_F,$\\
       $\rho_B^2\rho_F,  \rho_B^2\rho_F,  \rho_B^2\rho_F,  \rho_B^2\rho_F\}$ & $\rho_B^2\rho_F,  \rho_B^2\rho_F,  \rho_B^2\rho_F,  \rho_B^2\rho_F\}$
       \\\hline
         ${\bm B}_L$ eigenvalues (dropping all dashings)   &   ${\bm B}_R$ eigenvalues (dropping all dashings) \\\hline\hline
        $\{  -\rho_B^2\rho_F, -\rho_B^2\rho_F, -\rho_B^2\rho_F, -\rho_B^2\rho_F, $ &  $\{  -\rho_B^2\rho_F, -\rho_B^2\rho_F, -\rho_B^2\rho_F, -\rho_B^2\rho_F,$\\
       $\rho_B^2\rho_F,  \rho_B^2\rho_F,  \rho_B^2\rho_F,  \rho_B^2\rho_F\}$ & $\rho_B^2\rho_F,  \rho_B^2\rho_F,  \rho_B^2\rho_F,  \rho_B^2\rho_F\}$
       \\\hline
    \end{tabular}
    \caption{Results of ${\bm B}_L$ and ${\bm B}_R$ eigenvalues with/without dashings for the adinkra in Fig \ref{fig:adink48}}
    \label{tab:result_GR(8,4)}
\end{table}

Consider the ${\bm B}{}^2$ matrix, which is 
\begin{equation}
    {\bm B}{}^2 ~=~ \left[
\begin{matrix}
{\bm B}_L^2 & 0 \\
0 & {\bm B}_R^2  \\
\end{matrix} \right]
\end{equation}

Eigenvalues for ${\bm B}_L^2$ matrix (which is diagonal) are
\begin{equation}
\label{equ:d8N4BL2}
\begin{split}
    \{ & \rho_B^4\rho_F^2, \rho_B^4\rho_F^2, \rho_B^4\rho_F^2, \rho_B^4\rho_F^2, \rho_B^4\rho_F^2,  \rho_B^4\rho_F^2,  \rho_B^4\rho_F^2,  \rho_B^4\rho_F^2\}
    \end{split}
\end{equation}

Eigenvalues for ${\bm B}_R^2$ matrix (which is diagonal) are
\begin{equation}
\begin{split}
   \{ & \rho_B^4\rho_F^2, \rho_B^4\rho_F^2, \rho_B^4\rho_F^2, \rho_B^4\rho_F^2, \rho_B^4\rho_F^2,  \rho_B^4\rho_F^2,  \rho_B^4\rho_F^2,  \rho_B^4\rho_F^2\}
    \end{split}
\end{equation}

Then we can find: 
\begin{align}
    \det({\bm B}_L^2) ~=~ \rho_B^{32}\rho_F^{16} \\
    \det({\bm B}_R^2) ~=~ \rho_B^{32}\rho_F^{16} \label{equ:d8N4detBR2}
\end{align}
which are consistent with our conjecture: 
\begin{equation}
    \det({\bm B}_L^2) ~=~ \det({\bm B}_R^2) ~=~ \rho_B^{N\times\text{(\# of bosons lifted )}}\rho_F^{N\times\text{(\# of fermions lifted )}}
    \label{equ:conjecture}
\end{equation}

\newpage
\subsection{Step-by-step Results}
\begin{figure}[htp!]
    \centering
    \includegraphics[width=0.5\textwidth]{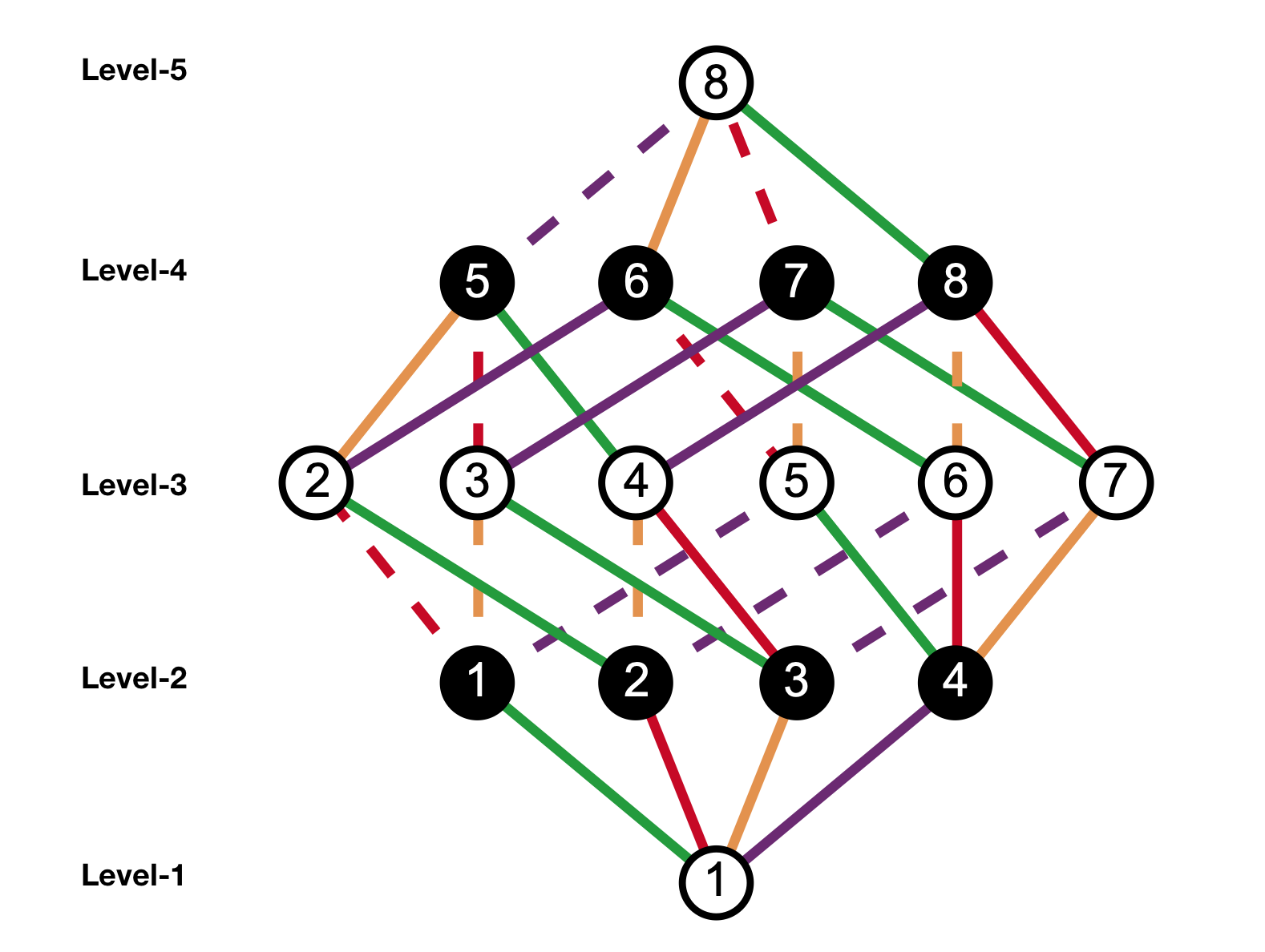}
    \caption{An Example of 4-color Non-minimal Adinkra}
    \label{fig:adink48_level}
\end{figure}
Look at Figure~\ref{fig:adink48_level}, start the analysis from the valise adinkra in which all bosons live in level-1 and all fermions live in level-2. In order to get the dimond adinkra as Figure~\ref{fig:adink48_level} shown, there are three steps to do the transformations. 

Note that lifting or lowering nodes would not influence dashing properties, so in all following steps, eigenvalues after dropping all dashing are the same as with dashing. 

\begin{figure}[htp!]
    \centering
    \includegraphics[width=0.5\textwidth]{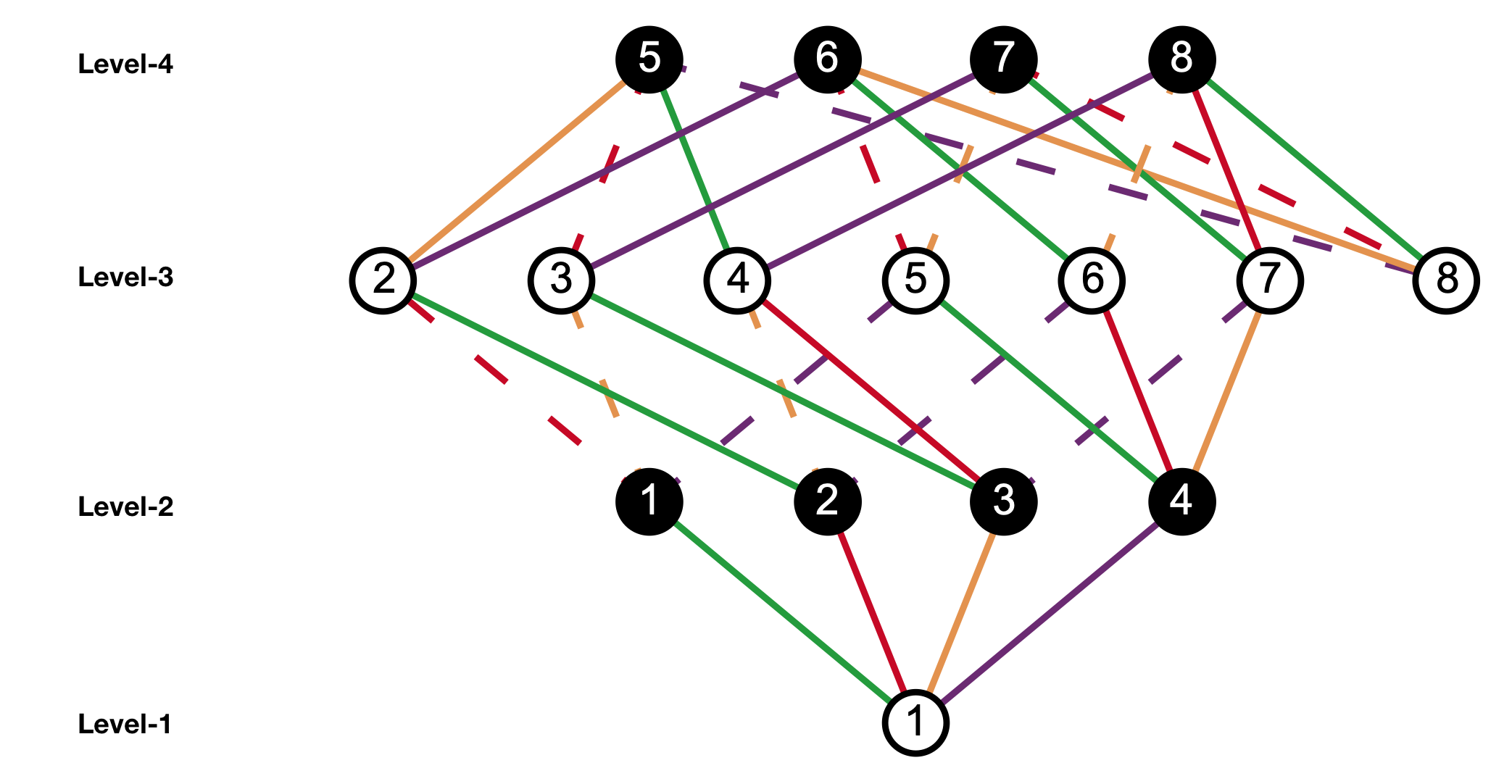}
    \caption{An Example of 4-color $<1|4|7|4>$ Non-minimal Adinkra}
    \label{fig:adink1474_level}
\end{figure}

In the following tables, we use $<|>$ notation to denote the ``shape" of adinkras we met in intermediate steps. The number sitting next to $<$ is the number of bosons living in level-1. $|$ is to divide adjacent levels. Thus the second number is the number of fermions living in level-2, and so on. 
So $<8|8>$ is the valise adinkra, our starting point, and the one in Figure \ref{fig:adink48_level} is $<1|4|6|4|1>$.
Figure \ref{fig:adink1474_level} shows the $<1|4|7|4>$ type adinkra.

\textbf{Step 1: Raise Boson 2-8 to Level-3.} See Table \ref{tab:d8N4Step1}. All values in the last column of Table \ref{tab:d8N4Step1} satisfy our conjecture (\ref{equ:conjecture}).

\begin{table}[htp!]
\small
\centering
\begin{tabular}{|c|c|c|c|}
\hline
  & ${\bm B}_{L}$ eigenvalues   &    ${\bm B}_{R}$ eigenvalues  & $ \det({\bm B}_{L/R}^2)$ \\\hline\hline
\multirow{2}{*}{$<8|8>$} & 
$\{   \, -1 \, ,  - 1 \, , - 1 \, ,  -1 \, ,$ & 
$\{   \, -1 \, ,  - 1 \, , - 1 \, ,  -1 \, , $ & \multirow{2}{*}{$\rho_B^{4\times0}$} \\
& $ 1 \, , 1 \, ,  1 \, ,  1 \, \}$ & $1 \, , 1 \, ,  1 \, ,  1 \, \}$ & \\\hline
\multirow{2}{*}{$<7|8|1>$} & 
$\{  \, -1 \, , -1 \, ,  1 \, , 1 \,,  $  &   
$\{  \, -1 \, , -1 \, ,  1 \, , 1 \,,  $  & \multirow{2}{*}{$\rho_B^{4\times1}$} \\
& $-\rho_B^{1/2} \, ,  -\rho_B^{1/2} \, ,   \rho_B^{1/2} \, , \rho_B^{1/2}\,  \}$ & $ -\rho_B^{1/2} \, ,  -\rho_B^{1/2} \, ,   \rho_B^{1/2} \, , \rho_B^{1/2}\,  \}$ & \\\hline
\multirow{2}{*}{$<6|8|2>$} & 
$\{    -\rho_B^{1/2}  , -\rho_B^{1/2}  ,   -\rho_B^{1/2}  ,  -\rho_B^{1/2} ,   $ & 
$\{   -1  , -1  ,  1  , 1 ,  $  & \multirow{2}{*}{$\rho_B^{4\times2}$} \\
& $\rho_B^{1/2}  ,\rho_B^{1/2}  ,  \rho_B^{1/2}  , \rho_B^{1/2}   \}$ & $ -\rho_B  , -\rho_B  ,  \rho_B  , \rho_B   \}$ & \\\hline 
\multirow{2}{*}{$<5|8|3>$} & 
$\{  -\rho_B^{1/2} , -\rho_B^{1/2}  ,   \rho_B^{1/2}  ,  \rho_B^{1/2} ,   $ & 
$\{  -\rho_B^{1/2} , -\rho_B^{1/2}  ,   \rho_B^{1/2}  ,  \rho_B^{1/2} ,   $   & \multirow{2}{*}{$\rho_B^{4\times3}$} \\
& $-\rho_B  , -\rho_B  ,  \rho_B  , \rho_B  \}$ & $-\rho_B  , -\rho_B  ,  \rho_B  , \rho_B  \}$ & \\\hline 
\multirow{2}{*}{$<4|8|4>$} & 
$\{ -\rho_B  , -\rho_B  ,  -\rho_B  , -\rho_B ,  $ & 
$\{ -\rho_B  , -\rho_B  ,  -\rho_B  , -\rho_B ,  $  & \multirow{2}{*}{$\rho_B^{4\times4}$} \\
& $\rho_B  , \rho_B  , \rho_B  , \rho_B   \}$ & $\rho_B  , \rho_B  , \rho_B  , \rho_B   \}$ & \\\hline 
\multirow{2}{*}{$<3|8|5>$} & 
$\{  -\rho_B  , -\rho_B ,  \rho_B  , \rho_B, $ & 
$\{  -\rho_B  , -\rho_B ,  \rho_B  , \rho_B, $  & \multirow{2}{*}{$\rho_B^{4\times5}$} \\
 & $ -\rho_B^{3/2},  -\rho_B^{3/2},  \rho_B^{3/2},  \rho_B^{3/2} \}$ & $ -\rho_B^{3/2},  -\rho_B^{3/2},  \rho_B^{3/2},  \rho_B^{3/2} \}$ &\\\hline 
\multirow{2}{*}{$<2|8|6>$} & 
$\{  -\rho_B^{3/2},  -\rho_B^{3/2} ,  -\rho_B^{3/2},  -\rho_B^{3/2},  \}$ & 
$\{  -\rho_B  , -\rho_B ,  \rho_B  , \rho_B ,   $  & \multirow{2}{*}{$\rho_B^{4\times6}$} \\
& $ \rho_B^{3/2},  \rho_B^{3/2}  , \rho_B^{3/2},  \rho_B^{3/2} $ & $-\rho_B^2  , -\rho_B^2 ,  \rho_B^2  , \rho_B^2   \}$ & \\\hline 
\multirow{2}{*}{$<1|8|7>$} & 
$\{  -\rho_B^{3/2},  -\rho_B^{3/2},  \rho_B^{3/2},  \rho_B^{3/2},$ & 
$\{    -\rho_B^{3/2},  -\rho_B^{3/2},  \rho_B^{3/2},  \rho_B^{3/2}, $ & \multirow{2}{*}{$\rho_B^{4\times7}$} \\
& $ -\rho_B^2, -\rho_B^2, \rho_B^2, \rho_B^2  \}$ & $-\rho_B^2, -\rho_B^2, \rho_B^2, \rho_B^2\}$ & \\\hline 
\end{tabular}
\caption{Eigenvalues of ${\bm B}_L$ and ${\bm B}_R$ matrices and determinants of ${\bm B}^2_L$ and ${\bm B}^2_R$ matrices in Step 1}
\label{tab:d8N4Step1}
\end{table}

\textbf{Step 2: Raise Fermions 5-8 to Level-4.} See Table \ref{tab:d8N4Step2}.  All values in the last column of Table \ref{tab:d8N4Step2} satisfy our conjecture (\ref{equ:conjecture}).

\begin{table}[htp!]
\small
\centering
\begin{tabular}{|c|c|c|}
\hline
  & ${\bm B}_{L}$ and ${\bm B}_{R}$ eigenvalues  & $ \det({\bm B}_{L/R}^2)$   \\\hline\hline
\multirow{2}{*}{$<1|7|7|1>$} &  ${\bm B}_{L}$: $\{ -\rho_B^{3/2},  \rho_B^{3/2},  -\rho_B^{2},  \rho_B^{2},  -\rho_B^{3/2}\rho_F^{1/2},  \rho_B^{3/2}\rho_F^{1/2} ,-\rho_B^{2}\rho_F^{1/2},  \rho_B^{2}\rho_F^{1/2} \}$  & \multirow{2}{*}{$\rho_B^{4\times7}\rho_F^{4\times1} $ } \\
 & ${\bm B}_{R}$: $\{ -\rho_B^{3/2},  \rho_B^{3/2},  -\rho_B^{2},  \rho_B^{2},  -\rho_B^{3/2}\rho_F^{1/2},  \rho_B^{3/2}\rho_F^{1/2} ,-\rho_B^{2}\rho_F^{1/2},  \rho_B^{2}\rho_F^{1/2} \}$ & \\  \hline
\multirow{2}{*}{$<1|6|7|2>$} & ${\bm B}_{L}$: $\{ -\rho_B^{3/2},  \rho_B^{3/2},  -\rho_B^{2},  \rho_B^{2},  -\rho_B^{3/2}\rho_F,  \rho_B^{3/2}\rho_F ,-\rho_B^{2}\rho_F,  \rho_B^{2}\rho_F \}$  & \multirow{2}{*}{$\rho_B^{4\times7}\rho_F^{4\times2} $ }\\
 & ${\bm B}_{R}$: $\{ -\rho_B^{3/2}\rho_F^{1/2},  -\rho_B^{3/2}\rho_F^{1/2}, \rho_B^{3/2}\rho_F^{1/2},  \rho_B^{3/2}\rho_F^{1/2} , -\rho_B^{2}\rho_F^{1/2},  \rho_B^{2}\rho_F^{1/2},  -\rho_B^{2}\rho_F^{1/2},  \rho_B^{2}\rho_F^{1/2} \}$ & \\  \hline
\multirow{2}{*}{$<1|5|7|3>$} & ${\bm B}_{L}$: $\{ -\rho_B^{3/2}\rho_F^{1/2},  \rho_B^{3/2}\rho_F^{1/2},  -\rho_B^{2}\rho_F^{1/2},  \rho_B^{2}\rho_F^{1/2},  -\rho_B^{3/2}\rho_F,  \rho_B^{3/2}\rho_F ,-\rho_B^{2}\rho_F,  \rho_B^{2}\rho_F \}$  & \multirow{2}{*}{$\rho_B^{4\times7}\rho_F^{4\times3} $ }\\
 & ${\bm B}_{R}$: $\{ -\rho_B^{3/2}\rho_F^{1/2},  \rho_B^{3/2}\rho_F^{1/2},  -\rho_B^{2}\rho_F^{1/2},  \rho_B^{2}\rho_F^{1/2},  -\rho_B^{3/2}\rho_F,  \rho_B^{3/2}\rho_F ,-\rho_B^{2}\rho_F,  \rho_B^{2}\rho_F \}$ & \\  \hline
\multirow{2}{*}{$<1|4|7|4>$} & ${\bm B}_{L}$: $\{  -\rho_B^{3/2}\rho_F, -\rho_B^{3/2}\rho_F,  \rho_B^{3/2}\rho_F,  \rho_B^{3/2}\rho_F, -\rho_B^2\rho_F, -\rho_B^2\rho_F, \rho_B^2\rho_F, \rho_B^2\rho_F \}$  & \multirow{2}{*}{$\rho_B^{4\times7}\rho_F^{4\times4} $ }\\
 &  ${\bm B}_{R}$: $\{  -\rho_B^{3/2}\rho_F, -\rho_B^{3/2}\rho_F,  \rho_B^{3/2}\rho_F,  \rho_B^{3/2}\rho_F, -\rho_B^2\rho_F, -\rho_B^2\rho_F, \rho_B^2\rho_F, \rho_B^2\rho_F \}$ &  \\  \hline
\end{tabular}
\caption{Eigenvalues of ${\bm B}_L$ and ${\bm B}_{R}$ matrices and determinants of ${\bm B}^2_L$ and ${\bm B}^2_R$ matrices in Step 2}
\label{tab:d8N4Step2}
\end{table}

\textbf{Step 3: Raise Boson 8 to Level-5} and we get the final results as Table \ref{tab:result_GR(8,4)} and Equation (\ref{equ:d8N4BL2}) to (\ref{equ:d8N4detBR2}).

\newpage
\section{Two Five-color Adinkras}
\label{sec:2-5CLR}

At this point, let us emphasize a matter of some importance that occurs whenever adinkras with
an odd number of colors is under consideration.  From (\ref{eq:BLBRdef}), it follows that the
index structure of the ${\bm B}_{R}$ and ${\bm B}_{L}$ matrices is very different.  Explicitly
expressions describing the matrix entries for each, take the forms
\be {
{\bm B}_{L} ~=~ \left( {B}_{L}\right){}_{j \, {k}}
~~~,~~~ {\bm B}_{R} ~=~ \left( {B}_{R}\right){}_{{\hat j} \, {\hat k}} ~~~,
} \label{eq:BLBRe}\ee
if $N$ is even but also
\be {
{\bm B}_{L} ~=~ \left( {B}_{L}\right){}_{j \, {\hat k}}
~~~,~~~ {\bm B}_{R} ~=~ \left( {B}_{R}\right){}_{{\hat j} \, {k}} ~~~,
} \label{eq:BLBRo}\ee
if $N$ is odd. This observation has some powerful implications for the multiplications of the 
the ${\bm B}_{R}$ and ${\bm B}_{L}$ matrices:
\newline \indent 
(a.) if $N$ is even, only eigenvalues of $({\bm B}_{R}){}^p$ and $({\bm B}_{L}){}^{q}$ for any
real numbers \newline $~~~~~~~~~~~$ $p$ and $q$ have well defined mathematical meanings, and 
\newline \indent 
(b.) if $N$ is odd, only eigenvalues of $({\bm B}_{R} \, {\bm B}_{L}){}^p$ and $({\bm B}_{L}\,
{\bm B}_R)^{q}$ for any real \newline  $~~~~~~~~~~~$ numbers $p$ and $q$ have well defined 
mathematical meanings.

\begin{figure}[htp!]
    \centering
    \includegraphics[width=0.9\textwidth]{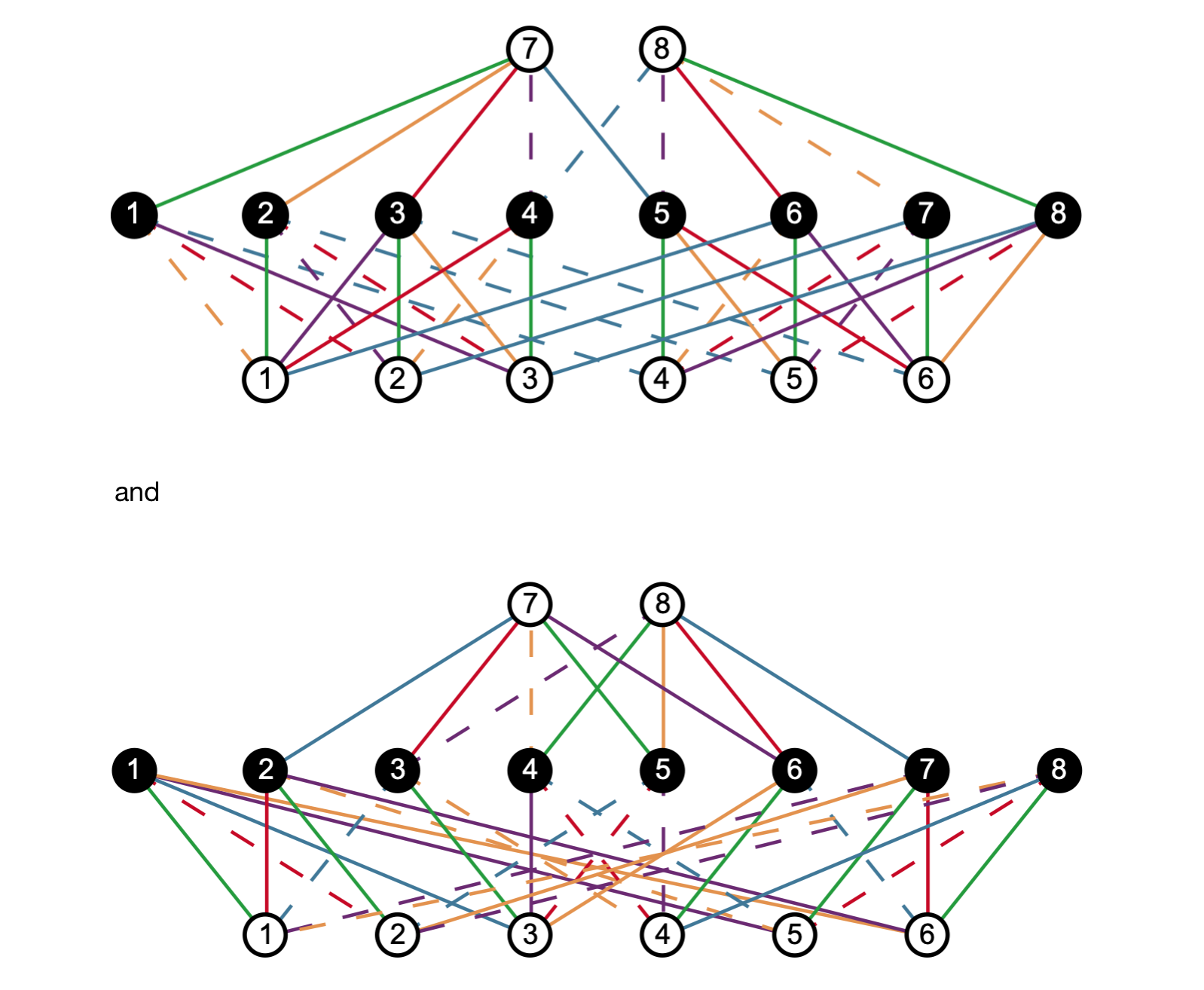}
    \caption{Two Five-color Adinkras}
    \label{fig:N5adink}
\end{figure}

In Fig.\ \ref{fig:N5adink}, there are shown two five-color adinkras.

Since in this case, $N = 5$ is odd, we have to study the eigenvalues of Banchoff ${\bm B}_R {\bm B}_L$ and ${\bm B}_L {\bm B}_R$
matrices. For each of these, we respectively find
\be{ \eqalign{
{\bm B}_R {\bm B}_L ~&=~ {\bm  {\rm R}}{}_{{}_{5}}{\bm  {\rm L}}{}_{{}_{4}}{\bm  {\rm R}}{}_{{}_{3}}{\bm  {\rm L}}{}_{{}_{2}}{\bm  {\rm R}}{}_{{}_{1}}{\bm  {\rm L}}{}_{{}_{5}}{\bm  {\rm R}}{}_{{}_{4}}{\bm  {\rm L}}{}_{{}_{3}}{\bm  {\rm R}}{}_{{}_{2}}{\bm  {\rm L}}{}_{{}_{1}}   ~~~~, \cr
{\bm B}_L {\bm B}_R ~&=~    {\bm  {\rm L}}{}_{{}_{5}}{\bm  {\rm R}}{}_{{}_{4}}{\bm  {\rm L}}{}_{{}_{3}}{\bm  {\rm R}}{}_{{}_{2}}{\bm  {\rm L}}{}_{{}_{1}} 
{\bm  {\rm R}}{}_{{}_{5}}{\bm  {\rm L}}{}_{{}_{4}}{\bm  {\rm R}}{}_{{}_{3}}{\bm  {\rm L}}{}_{{}_{2}}{\bm  {\rm R}}{}_{{}_{1}}
~~~~. \cr
}}\ee

For the first adinkra, the ${\bm {\rm L}}$-matrices are:
\begin{align}
    {\bm  {\rm L}}{}_{{}_{1}} ~&=~ \left(
\begin{matrix}
0 & 1 & 0 & 0 & 0 & 0 & 0 & 0\\
0 & 0 & 1 & 0 & 0 & 0 & 0 & 0\\
0 & 0 & 0 & 1 & 0 & 0 & 0 & 0\\
0 & 0 & 0 & 0 & 1 & 0 & 0 & 0\\
0 & 0 & 0 & 0 & 0 & 1 & 0 & 0  \\
0 & 0 & 0 & 0 & 0 & 0 & 1 & 0\\
1 & 0 & 0 & 0 & 0 & 0 & 0 & 0\\
0 & 0 & 0 & 0 & 0 & 0 & 0 & 1\\
\end{matrix}
\right) ~~~~~~,~~~~~~
    {\bm  {\rm L}}{}_{{}_{2}} ~=~ \left(
\begin{matrix}
0 & 0 & 1 & 0 & 0 & 0 & 0 & 0\\
0 & -1 & 0 & 0 & 0 & 0 & 0 & 0\\
1 & 0 & 0 & 0 & 0 & 0 & 0 & 0\\
0 & 0 & 0 & 0 & 0 & 0 & 0 & 1\\
0 & 0 & 0 & 0 & 0 & 0 & -1 & 0\\
0 & 0 & 0 & 0 & 0 & 1 & 0 & 0\\
0 & 0 & 0 & -1 & 0 & 0 & 0 & 0\\
0 & 0 & 0 & 0 & -1 & 0 & 0 & 0\\
\end{matrix}
\right) ~~, \\
{\bm  {\rm L}}{}_{{}_{3}} ~&=~ \left(
\begin{matrix}
-1 & 0 & 0 & 0 & 0 & 0 & 0 & 0\\
0 & 0 & 0 & -1 & 0 & 0 & 0 & 0\\
0 & 0 & 1 & 0 & 0 & 0 & 0 & 0\\
0 & 0 & 0 & 0 & 0 & -1 & 0 & 0\\
0 & 0 & 0 & 0 & 1 & 0 & 0 & 0\\
0 & 0 & 0 & 0 & 0 & 0 & 0 & 1\\
0 & 1 & 0 & 0 & 0 & 0 & 0 & 0\\
0 & 0 & 0 & 0 & 0 & 0 & -1 & 0\\
\end{matrix}
\right) ~~, 
    {\bm  {\rm L}}{}_{{}_{4}} ~=~ \left(
\begin{matrix}
0 & 0 & 0 & 1 & 0 & 0 & 0 & 0\\
-1 & 0 & 0 & 0 & 0 & 0 & 0 & 0\\
0 & -1 & 0 & 0 & 0 & 0 & 0 & 0\\
0 & 0 & 0 & 0 & 0 & 0 & -1 & 0\\
0 & 0 & 0 & 0 & 0 & 0 & 0 & -1\\
0 & 0 & 0 & 0 & 1 & 0 & 0 & 0\\
0 & 0 & 1 & 0 & 0 & 0 & 0 & 0\\
0 & 0 & 0 & 0 & 0 & 1 & 0 & 0\\
\end{matrix}
\right) ~~, \\
{\bm {\rm L}}_5 ~&=~ \left(
\begin{matrix}
0 & 0 & 0 & 0 & 0 & 1 & 0 & 0\\
0 & 0 & 0 & 0 & 0 & 0 & 1 & 0\\
0 & 0 & 0 & 0 & 0 & 0 & 0 & 1\\
-1 & 0 & 0 & 0 & 0 & 0 & 0 & 0\\
0 & -1 & 0 & 0 & 0 & 0 & 0 & 0\\
0 & 0 & -1 & 0 & 0 & 0 & 0 & 0\\
0 & 0 & 0 & 0 & 1 & 0 & 0 & 0\\
0 & 0 & 0 & -1 & 0 & 0 & 0 & 0\\
\end{matrix}
\right) ~~.
\end{align}

By analyzing the HYMNs, we obtain the results listed in Table \ref{tab:result_N51}. 
\begin{table}[htp!]
    \centering
    \begin{tabular}{|c|c|}
    \hline
    ${\bm B}_L{\bm B}_R$ eigenvalues   &   ${\bm B}_R{\bm B}_L$ eigenvalues  \\\hline\hline
       $\{  1, 1, \rho_B  , \rho_B , \rho_B^2 , \rho_B^2 , \rho_B^2 , \rho_B^2 \} $ &  $\{  1, 1, \rho_B  , \rho_B , \rho_B^2 , \rho_B^2 , \rho_B^2 , \rho_B^2 \} $\\\hline
         ${\bm B}_L{\bm B}_R$ eigenvalues (dropping all dashings)   &   ${\bm B}_R{\bm B}_L$ eigenvalues (dropping all dashings) \\\hline\hline
        $\{  1, 1, \rho_B  , \rho_B , \rho_B^2 , \rho_B^2 , \rho_B^2 , \rho_B^2 \} $ &  $\{  1, 1, \rho_B  , \rho_B , \rho_B^2 , \rho_B^2 , \rho_B^2 , \rho_B^2 \} $
       \\\hline
    \end{tabular}
    \caption{Results of ${\bm B}_L {\bm B}_R$ and ${\bm B}_R {\bm B}_L$ eigenvalues with/without dashings for the first adinkra \newline
    $ ~~~~~~~~~\,~~~~~~$ in Figure \ref{fig:N5adink}}
    \label{tab:result_N51}
\end{table}

For the second adinkra, the ${\bm {\rm L}}$-matrices are:
\begin{align}
    {\bm  {\rm L}}{}_{{}_{1}} ~&=~ \left(
\begin{matrix}
1 & 0 & 0 & 0 & 0 & 0 & 0 & 0\\
0 & 1 & 0 & 0 & 0 & 0 & 0 & 0\\
0 & 0 & 1 & 0 & 0 & 0 & 0 & 0\\
0 & 0 & 0 & 0 & 0 & 1 & 0 & 0\\
0 & 0 & 0 & 0 & 0 & 0 & 1 & 0\\
0 & 0 & 0 & 0 & 0 & 0 & 0 & 1\\
0 & 0 & 0 & 0 & 1 & 0 & 0 & 0\\
0 & 0 & 0 & 1 & 0 & 0 & 0 & 0\\
\end{matrix}
\right) ~~~~~~,~~~~~~
    {\bm  {\rm L}}{}_{{}_{2}}~=~ \left(
\begin{matrix}
0 & 0 & 0 & 0 & 0 & 0 & -1 & 0\\
0 & 0 & 0 & 0 & 0 & 0 & 0 & -1\\
0 & 0 & 0 & 1 & 0 & 0 & 0 & 0\\
0 & 0 & 0 & 0 & -1 & 0 & 0 & 0\\
1 & 0 & 0 & 0 & 0 & 0 & 0 & 0\\
0 & 1 & 0 & 0 & 0 & 0 & 0 & 0 \\
0 & 0 & 0 & 0 & 0 & 1 & 0 & 0\\
0 & 0 & -1 & 0 & 0 & 0 & 0 & 0\\
\end{matrix}
\right) ~~, \\
{\bm  {\rm L}}{}_{{}_{3}} ~&=~ \left(
\begin{matrix}
0 & 0 & 0 & 0 & 0 & 0 & 0 & -1\\
0 & 0 & 0 & 0 & 0 & 0 & 1 & 0\\
0 & 0 & 0 & 0 & 0 & 1 & 0 & 0\\
0 & 0 & -1 & 0 & 0 & 0 & 0 & 0\\
0 & -1 & 0 & 0 & 0 & 0 & 0 & 0 \\
1 & 0 & 0 & 0 & 0 & 0 & 0 & 0  \\
0 & 0 & 0 & -1 & 0 & 0 & 0 & 0\\
0 & 0 & 0 & 0 & 1 & 0 & 0 & 0\\
\end{matrix}
\right) ~~, 
    {\bm  {\rm L}}{}_{{}_{4}} ~=~ \left(
\begin{matrix}
0 & 1 & 0 & 0 & 0 & 0 & 0 & 0\\
-1 & 0 & 0 & 0 & 0 & 0 & 0 & 0\\
0 & 0 & 0 & 0 & -1 & 0 & 0 & 0\\
0 & 0 & 0 & -1 & 0 & 0 & 0 & 0\\
0 & 0 & 0 & 0 & 0 & 0 & 0 & -1\\
0 & 0 & 0 & 0 & 0 & 0 & 1 & 0\\
0 & 0 & 1 & 0 & 0 & 0 & 0 & 0\\
0 & 0 & 0 & 0 & 0 & 1 & 0 & 0\\
\end{matrix}
\right) ~~, \\
{\bm  {\rm L}}{}_{{}_{5}} ~&=~ \left(
\begin{matrix}
0 & 0 & -1 & 0 & 0 & 0 & 0 & 0\\
0 & 0 & 0 & 0 & -1 & 0 & 0 & 0\\
1 & 0 & 0 & 0 & 0 & 0 & 0 & 0\\
0 & 0 & 0 & 0 & 0 & 0 & 0 & 1\\
0 & 0 & 0 & -1 & 0 & 0 & 0 & 0\\
0 & 0 & 0 & 0 & 0 & -1 & 0 & 0 \\
0 & 1 & 0 & 0 & 0 & 0 & 0 & 0\\
0 & 0 & 0 & 0 & 0 & 0 & 1 & 0\\
\end{matrix}
\right) ~~.
\end{align}

By analyzing the HYMNs for this case we find the results listed in Table \ref{tab:result_N52}. 
\begin{table}[htp!]
    \centering
    \begin{tabular}{|c|c|}
    \hline
    ${\bm B}_L{\bm B}_R$ eigenvalues   &   ${\bm B}_R{\bm B}_L$ eigenvalues  \\\hline\hline
       $\{  \rho_B , \rho_B , \rho_B , \rho_B , \rho_B , \rho_B , \rho_B^2 , \rho_B^2 \} $ &  $\{  \rho_B , \rho_B , \rho_B , \rho_B , \rho_B , \rho_B , \rho_B^2 , \rho_B^2 \}$\\\hline
         ${\bm B}_L{\bm B}_R$ eigenvalues (dropping all dashings)   &   ${\bm B}_R{\bm B}_L$ eigenvalues (dropping all dashings) \\\hline\hline
        $\{  \rho_B , \rho_B , \rho_B , \rho_B , \rho_B , \rho_B , \rho_B^2 , \rho_B^2 \} $ &  $\{  \rho_B , \rho_B , \rho_B , \rho_B , \rho_B , \rho_B , \rho_B^2 , \rho_B^2 \}$
       \\\hline
    \end{tabular}
    \caption{Results of ${\bm B}_L$ and ${\bm B}_R$ eigenvalues with/without dashings for the second adinkra in Figure \ref{fig:N5adink}}
    \label{tab:result_N52}
\end{table}

\newpage
\section{Two Six-color Adinkras\label{sec:2-6color}}

\begin{figure}[htp!]
    \centering
    \includegraphics[width=0.9\textwidth]{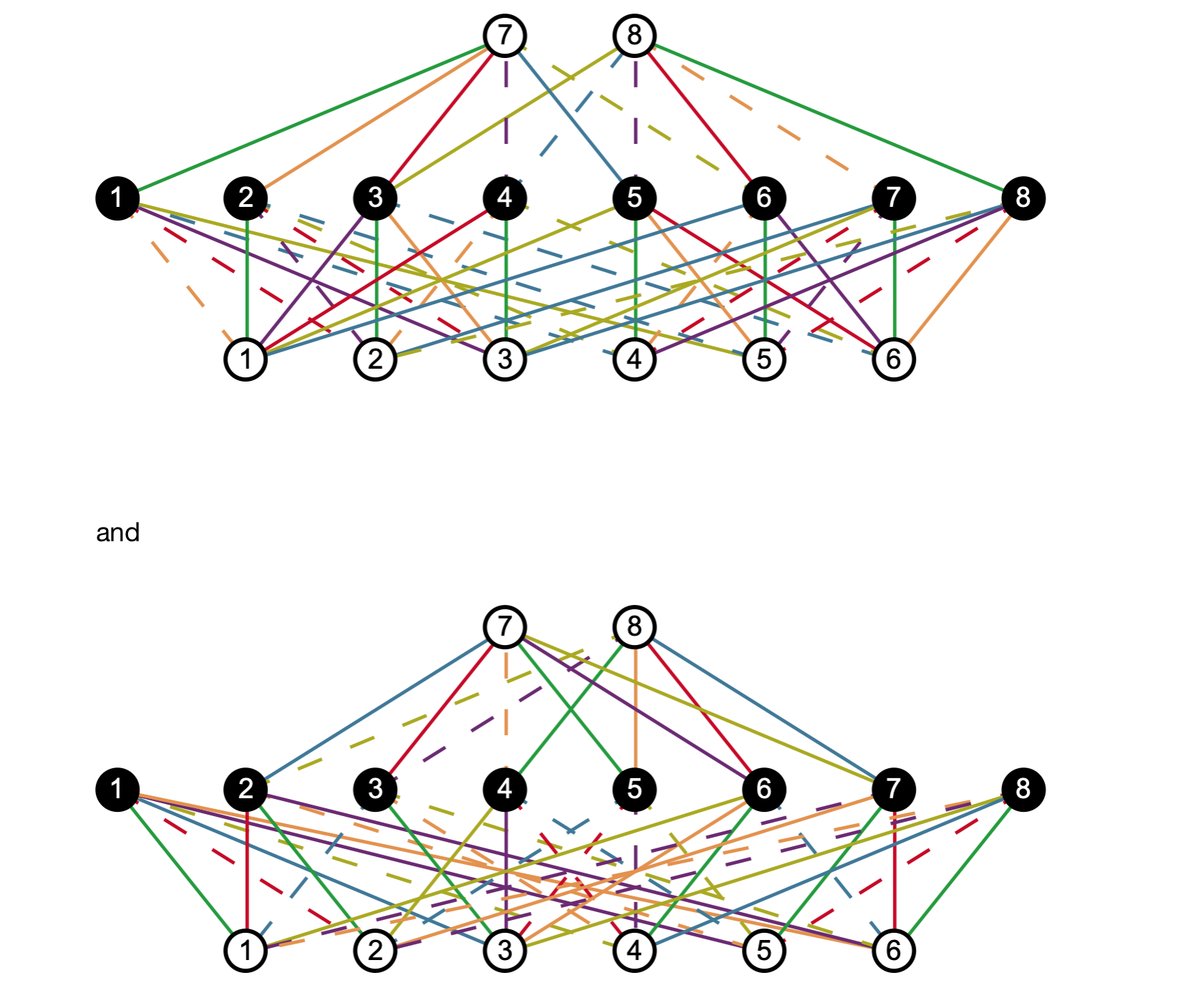}
    \caption{Two Six-color Adinkras}
    \label{fig:N6adink}
\end{figure}

In this chapter, we are going to study HYMNs for six-color adinkras. In Figure \ref{fig:N6adink}, there are shown two six-color adinkras. 

Since in this case, $N=6$ is even, we define Banchoff $\bm B$-matrices in the way similar to $N=4$ case. 
\begin{align}
    {\bm B}_L ~=&~ {\bm  {\rm L}}{}_{{}_{6}}{\bm  {\rm R}}{}_{{}_{5}}{\bm  {\rm L}}{}_{{}_{4}}{\bm  {\rm R}}{}_{{}_{3}}{\bm  {\rm L}}{}_{{}_{2}}{\bm  {\rm R}}{}_{{}_{1}}~~~,\\
    {\bm B}_R ~=&~ {\bm  {\rm R}}{}_{{}_{6}}{\bm  {\rm L}}{}_{{}_{5}}{\bm  {\rm R}}{}_{{}_{4}}{\bm  {\rm L}}{}_{{}_{3}}{\bm  {\rm R}}{}_{{}_{2}}{\bm  {\rm L}}{}_{{}_{1}}~~~.
\end{align}

For the first adinkra, the ${\bm {\rm L}}$-matrices are:
\begin{align}
    {\bm  {\rm L}}{}_{{}_{1}} ~&=~ \left(
\begin{matrix}
0 & 1 & 0 & 0 & 0 & 0 & 0 & 0\\
0 & 0 & 1 & 0 & 0 & 0 & 0 & 0\\
0 & 0 & 0 & 1 & 0 & 0 & 0 & 0\\
0 & 0 & 0 & 0 & 1 & 0 & 0 & 0\\
0 & 0 & 0 & 0 & 0 & 1 & 0 & 0  \\
0 & 0 & 0 & 0 & 0 & 0 & 1 & 0\\
1 & 0 & 0 & 0 & 0 & 0 & 0 & 0\\
0 & 0 & 0 & 0 & 0 & 0 & 0 & 1\\
\end{matrix}
\right) ~~~~~~,~~~~~~
    {\bm  {\rm L}}{}_{{}_{2}} ~=~ \left(
\begin{matrix}
0 & 0 & 1 & 0 & 0 & 0 & 0 & 0\\
0 & -1 & 0 & 0 & 0 & 0 & 0 & 0\\
1 & 0 & 0 & 0 & 0 & 0 & 0 & 0\\
0 & 0 & 0 & 0 & 0 & 0 & 0 & 1\\
0 & 0 & 0 & 0 & 0 & 0 & -1 & 0\\
0 & 0 & 0 & 0 & 0 & 1 & 0 & 0\\
0 & 0 & 0 & -1 & 0 & 0 & 0 & 0\\
0 & 0 & 0 & 0 & -1 & 0 & 0 & 0\\
\end{matrix}
\right) ~~, \\
{\bm  {\rm L}}{}_{{}_{3}} ~&=~ \left(
\begin{matrix}
-1 & 0 & 0 & 0 & 0 & 0 & 0 & 0\\
0 & 0 & 0 & -1 & 0 & 0 & 0 & 0\\
0 & 0 & 1 & 0 & 0 & 0 & 0 & 0\\
0 & 0 & 0 & 0 & 0 & -1 & 0 & 0\\
0 & 0 & 0 & 0 & 1 & 0 & 0 & 0\\
0 & 0 & 0 & 0 & 0 & 0 & 0 & 1\\
0 & 1 & 0 & 0 & 0 & 0 & 0 & 0\\
0 & 0 & 0 & 0 & 0 & 0 & -1 & 0\\
\end{matrix}
\right) ~~, 
    {\bm  {\rm L}}{}_{{}_{4}} ~=~ \left(
\begin{matrix}
0 & 0 & 0 & 1 & 0 & 0 & 0 & 0\\
-1 & 0 & 0 & 0 & 0 & 0 & 0 & 0\\
0 & -1 & 0 & 0 & 0 & 0 & 0 & 0\\
0 & 0 & 0 & 0 & 0 & 0 & -1 & 0\\
0 & 0 & 0 & 0 & 0 & 0 & 0 & -1\\
0 & 0 & 0 & 0 & 1 & 0 & 0 & 0\\
0 & 0 & 1 & 0 & 0 & 0 & 0 & 0\\
0 & 0 & 0 & 0 & 0 & 1 & 0 & 0\\
\end{matrix}
\right) ~~, \\
{\bm  {\rm L}}{}_{{}_{5}} ~&=~ \left(
\begin{matrix}
0 & 0 & 0 & 0 & 0 & 1 & 0 & 0\\
0 & 0 & 0 & 0 & 0 & 0 & 1 & 0\\
0 & 0 & 0 & 0 & 0 & 0 & 0 & 1\\
-1 & 0 & 0 & 0 & 0 & 0 & 0 & 0\\
0 & -1 & 0 & 0 & 0 & 0 & 0 & 0\\
0 & 0 & -1 & 0 & 0 & 0 & 0 & 0\\
0 & 0 & 0 & 0 & 1 & 0 & 0 & 0\\
0 & 0 & 0 & -1 & 0 & 0 & 0 & 0\\
\end{matrix}
\right) ~~, 
{\bm  {\rm L}}{}_{{}_{6}} ~=~ \left(
\begin{matrix}
0 & 0 & 0 & 0 & 1 & 0 & 0 & 0\\
0 & 0 & 0 & 0 & 0 & 0 & 0 & -1\\
0 & 0 & 0 & 0 & 0 & 0 & 1 & 0\\
0 & -1 & 0 & 0 & 0 & 0 & 0 & 0\\
1 & 0 & 0 & 0 & 0 & 0 & 0 & 0\\
0 & 0 & 0 & -1 & 0 & 0 & 0 & 0\\
0 & 0 & 0 & 0 & 0 & -1 & 0 & 0\\
0 & 0 & 1 & 0 & 0 & 0 & 0 & 0\\
\end{matrix}
\right) ~~.
\end{align}

By analyzing the HYMNs, we get results listed in Table \ref{tab:result_N61}. 
\begin{table}[htp!]
    \centering
    \begin{tabular}{|c|c|}
    \hline
    ${\bm B}_L$ eigenvalues   &   ${\bm B}_R$ eigenvalues  \\\hline\hline
       $\{  -i\sqrt{\rho_B},
    -i\sqrt{\rho_B},
    i\sqrt{\rho_B},
     i\sqrt{\rho_B}, $ &  $\{  -i\sqrt{\rho_B},
     -i\sqrt{\rho_B},
     i\sqrt{\rho_B},
     i\sqrt{\rho_B},$\\
       $-i\rho_B,
     -i\rho_B,
     i\rho_B,
     i\rho_B\}$ & $-i\rho_B,
     -i\rho_B,
     i\rho_B,
     i\rho_B\}$
       \\\hline
         ${\bm B}_L$ eigenvalues (dropping all dashings)   &   ${\bm B}_R$ eigenvalues (dropping all dashings) \\\hline\hline
        $\{  -\sqrt{\rho_B},
    -\sqrt{\rho_B},
    \sqrt{\rho_B},
     \sqrt{\rho_B} $ &  $\{  -\sqrt{\rho_B},
     -\sqrt{\rho_B},
     \sqrt{\rho_B},
     \sqrt{\rho_B},$\\
       $ -\rho_B,
     -\rho_B,
     \rho_B,
     \rho_B\}$ & $-\rho_B,
     -\rho_B,
     \rho_B,
     \rho_B\}$
       \\\hline
    \end{tabular}
    \caption{Results of ${\bm B}_L$ and ${\bm B}_R$ eigenvalues with/without dashings for the first adinkra in Fig \ref{fig:N6adink}}
    \label{tab:result_N61}
\end{table}

For the second adinkra, the ${\bm {\rm L}}$-matrices are:
\begin{align}
    {\bm  {\rm L}}{}_{{}_{1}} ~&=~ \left(
\begin{matrix}
1 & 0 & 0 & 0 & 0 & 0 & 0 & 0\\
0 & 1 & 0 & 0 & 0 & 0 & 0 & 0\\
0 & 0 & 1 & 0 & 0 & 0 & 0 & 0\\
0 & 0 & 0 & 0 & 0 & 1 & 0 & 0\\
0 & 0 & 0 & 0 & 0 & 0 & 1 & 0\\
0 & 0 & 0 & 0 & 0 & 0 & 0 & 1\\
0 & 0 & 0 & 0 & 1 & 0 & 0 & 0\\
0 & 0 & 0 & 1 & 0 & 0 & 0 & 0\\
\end{matrix}
\right) ~~~~~~,~~~~~~
    {\bm  {\rm L}}{}_{{}_{2}}~=~ \left(
\begin{matrix}
0 & 0 & 0 & 0 & 0 & 0 & -1 & 0\\
0 & 0 & 0 & 0 & 0 & 0 & 0 & -1\\
0 & 0 & 0 & 1 & 0 & 0 & 0 & 0\\
0 & 0 & 0 & 0 & -1 & 0 & 0 & 0\\
1 & 0 & 0 & 0 & 0 & 0 & 0 & 0\\
0 & 1 & 0 & 0 & 0 & 0 & 0 & 0 \\
0 & 0 & 0 & 0 & 0 & 1 & 0 & 0\\
0 & 0 & -1 & 0 & 0 & 0 & 0 & 0\\
\end{matrix}
\right) ~~, \\
{\bm  {\rm L}}{}_{{}_{3}} ~&=~ \left(
\begin{matrix}
0 & 0 & 0 & 0 & 0 & 0 & 0 & -1\\
0 & 0 & 0 & 0 & 0 & 0 & 1 & 0\\
0 & 0 & 0 & 0 & 0 & 1 & 0 & 0\\
0 & 0 & -1 & 0 & 0 & 0 & 0 & 0\\
0 & -1 & 0 & 0 & 0 & 0 & 0 & 0 \\
1 & 0 & 0 & 0 & 0 & 0 & 0 & 0  \\
0 & 0 & 0 & -1 & 0 & 0 & 0 & 0\\
0 & 0 & 0 & 0 & 1 & 0 & 0 & 0\\
\end{matrix}
\right) ~~, 
    {\bm  {\rm L}}{}_{{}_{4}} ~=~ \left(
\begin{matrix}
0 & 1 & 0 & 0 & 0 & 0 & 0 & 0\\
-1 & 0 & 0 & 0 & 0 & 0 & 0 & 0\\
0 & 0 & 0 & 0 & -1 & 0 & 0 & 0\\
0 & 0 & 0 & -1 & 0 & 0 & 0 & 0\\
0 & 0 & 0 & 0 & 0 & 0 & 0 & -1\\
0 & 0 & 0 & 0 & 0 & 0 & 1 & 0\\
0 & 0 & 1 & 0 & 0 & 0 & 0 & 0\\
0 & 0 & 0 & 0 & 0 & 1 & 0 & 0\\
\end{matrix}
\right) ~~, \\
{\bm  {\rm L}}{}_{{}_{5}} ~&=~ \left(
\begin{matrix}
0 & 0 & -1 & 0 & 0 & 0 & 0 & 0\\
0 & 0 & 0 & 0 & -1 & 0 & 0 & 0\\
1 & 0 & 0 & 0 & 0 & 0 & 0 & 0\\
0 & 0 & 0 & 0 & 0 & 0 & 0 & 1\\
0 & 0 & 0 & -1 & 0 & 0 & 0 & 0\\
0 & 0 & 0 & 0 & 0 & -1 & 0 & 0 \\
0 & 1 & 0 & 0 & 0 & 0 & 0 & 0\\
0 & 0 & 0 & 0 & 0 & 0 & 1 & 0\\
\end{matrix}
\right) ~~, 
{\bm  {\rm L}}{}_{{}_{6}} ~=~ \left(
\begin{matrix}
0 & 0 & 0 & 0 & 0 & 1 & 0 & 0\\
0 & 0 & 0 & 1 & 0 & 0 & 0 & 0\\
0 & 0 & 0 & 0 & 0 & 0 & 0 & 1\\
-1 & 0 & 0 & 0 & 0 & 0 & 0 & 0\\
0 & 0 & 0 & 0 & -1 & 0 & 0 & 0\\
0 & 0 & -1 & 0 & 0 & 0 & 0 & 0\\
0 & 0 & 0 & 0 & 0 & 0 & 1 & 0\\
0 & -1 & 0 & 0 & 0 & 0 & 0 & 0\\
\end{matrix}
\right) ~~.
\end{align}

By analyzing the HYMNs, we get results listed in Table \ref{tab:result_N62}. 
\begin{table}[htp!]
    \centering
    \begin{tabular}{|c|c|}
    \hline
    ${\bm B}_L$ eigenvalues   &   ${\bm B}_R$ eigenvalues  \\\hline\hline
       $\{  -i, -i, i, i, $ &  $\{  -i, i, -i\rho_B, -i\rho_B,$\\
       $-i\rho_B, i\rho_B, -i\rho_B^2, i\rho_B^2\}$ & $-i\rho_B, i\rho_B, i\rho_B, i\rho_B \}$
       \\\hline
         ${\bm B}_L$ eigenvalues (dropping all dashings)   &   ${\bm B}_R$ eigenvalues (dropping all dashings) \\\hline\hline
        $\{  -1, -1, 1, 1, $ &  $\{  -1, 1, -\rho_B, -\rho_B,$\\
       $  -\rho_B, \rho_B, -\rho_B^2, \rho_B^2\}$ & $-\rho_B, \rho_B, \rho_B, \rho_B\}$
       \\\hline
    \end{tabular}
    \caption{Results of ${\bm B}_L$ and ${\bm B}_R$ eigenvalues with/without dashings for the second adinkra in Fig \ref{fig:N6adink}}
    \label{tab:result_N62}
\end{table}
      
\newpage 
\section{Conclusion}
In this paper, we have extended our previous discussions about using HYMNs (height-yielding matrix numbers)
which are the eigenvalues \cite{Iso} of functions of the adjacency matrices associated with the $\bm {\rm L}$-matrics
and $\bm {\rm R} $-matrices derived from adinkras.  The traces and determinants of the Banchoff matrices defined in this paper yield polynomials that
are sensitive of the ``shapes'' of the adinkras in all cases examined.  Further study will be required
to support the current speculation that these polynomials split adinkras into equivalent classes.   Even 
more intriguing is the possibility that these polynomials could play an important role in the concept of 
``SUSY Holography," as introduced in the work of \cite{ZZ}. These topics will be the subject of future explorations.

  \vspace{.05in}
 \begin{center}
\parbox{4in}{{\it ``The object of pure Physics is the unfolding of the laws $~~$ of the
intelligible world; the object of pure Mathematics $~~$ that of unfolding the laws of human intelligence.'' \\ ${~}$ 
 ${~}$ 
\\ ${~}$ }\,\,-\,\, J. \, J. \, Sylvester $~~~~~~~~~$}
 \parbox{4in}{
 $~~$}  
 \end{center}
		
\noindent
{\bf Acknowledgments}\\[.1in] \indent
This research is supported in part by the endowment of the Ford Foundation Professorship 
of Physics at Brown University and the Brown Theoretical Physics Center. Yangrui Hu would 
like to acknowledge her participation in the second annual Brown University Adinkra 
Math/Phys Hangout" during 2017. This work was also partially supported by the National Science 
Foundation grant PHY-1620074/09-68854, the endowment of the John S.~Toll Professorship, and 
the University of Maryland Center for String \& Particle Theory.  Additionally S.\ J.\ G. 
acknowledges the generous support of the Roth Professorship and the very congenial and generous 
hospitality of the Dartmouth College physics department in the period of this investigation.

\newpage
\appendix
\section{Transformation Laws of 4D, \texorpdfstring{$\mathcal{N}=1$}{N=1} Minimal Supermultiplets\label{appen:transf_law}}

Although a lot of literature about Adinkras has presented transformation laws of ten minimal off-shell 4D, $\mathcal{N}=1$ supermultiplets, for example chapter three in \cite{HYMN1}, there are some chronic typos. 
For clarity, we include the corrected transformation laws in this appendix. 
Note that the convention is $\e^{0123}=+1$ and $\eta^{\mu\n}=\eta_{\m\n}={\rm diag}(-1,1,1,1)$.

$ {Chiral~Supermultiplet: ~(A, \, B, \,  \psi_a , \, F, \, G)}$
\be
 \eqalign{
{~~~~} {\rm D}_a A ~&=~ \psi_a  ~~~~~~~~~~~~~~,~~~~
{\rm D}_a B ~=~ i \, (\g^5){}_a{}^b \, \psi_b  ~~~~~~~~~, \cr
{\rm D}_a \psi_b ~&=~ i\, (\g^\m){}_{a \,b}\,  \pa_\m A 
~-~  (\g^5\g^\m){}_{a \,b} \, \pa_\m B ~-~ i \, C_{a\, b} 
\, F  ~+~  (\g^5){}_{ a \, b} G  ~~, \cr
{\rm D}_a F ~&=~  (\g^\m){}_a{}^b \,  
\pa_{\m}\, \psi_b   
~~~,~~~ 
{\rm D}_a G ~=~ i \,(\g^5\g^\m){}_a{}^b \, \pa_\m \,  
\psi_b  ~~~,} \label{CM}
\ee 

$ {Hodge-Dual~ \#1~Chiral~Supermultiplet: ~(A, \, B, \,  \psi_a , \, {\rm f}_{\mu
 \, \nu \, \rho}, \, G)}$
\be
 \eqalign{
{~~~~} {\rm D}_a A ~&=~ \psi_a  ~~~~~~~~~~~~~~,~~~~
{\rm D}_a B ~=~ i \, (\g^5){}_a{}^b \, \psi_b  ~~~~~~~~~, \cr
{\rm D}_a \psi_b ~&=~ i\, (\g^\m){}_{a \,b}\,  \pa_\m A 
~-~  (\g^5\g^\m){}_{a \,b} \, \pa_\m B ~-~ i \,  \frac 1{3!} \, C_{a\, b} 
\, (\e{}^{\s}{}^{\m}{}^{\n}{}^{\rho} \, \pa_{\s} {\rm f}_{\m
 \, \n \, \rho})  ~+~  (\g^5){}_{ a \, b} G  ~~, \cr
{\rm D}_a {\rm f}_{\m \, \n \, \rho} ~&=~ -\,  (\g^\s){}_a{}^b \,  
\e{}_{\s}{}_{\m}{}_{\n}{}_{\rho} \, \psi_b   
~~~,~~~ 
{\rm D}_a G ~=~ i \,(\g^5\g^\m){}_a{}^b \, \pa_\m \,  
\psi_b  ~~~,} \label{QTd1}
\ee 
${Hodge-Dual~ \#2~Chiral~Supermultiplet: ~(A, \, B, \,  \psi_a , \, F, \, {\rm 
g}_{\m \, \n \, \rho})}$
\be
 \eqalign{
{~~~~} {\rm D}_a A ~&=~ \psi_a  ~~~~~~~~~~~~~~,~~~~
{\rm D}_a B ~=~ i \, (\g^5){}_a{}^b \, \psi_b  ~~~~~~~~~, \cr
{\rm D}_a \psi_b ~&=~ i\, (\g^\m){}_{a \,b}\,  \pa_\m A 
~-~  (\g^5\g^\m){}_{a \,b} \, \pa_\m B ~-~ i \, C_{a\, b} 
\,F  ~+~  \frac 1{3!} \, (\g^5){}_{ a \, b}
\, (\e{}^{\s}{}^{\m}{}^{\n}{}^{\rho} \, \pa_{\s} {\rm g}_{\m
 \, \n \, \rho})  ~~, \cr
{\rm D}_a F ~&=~  (\g^\m){}_a{}^b \, \pa_\m \, \psi_b   
~~~,~~~ 
{\rm D}_a {\rm g}_{\m \, \n \, \rho} ~=~ -\,  (\g^5 \g^\s){}_a{}^b \,  
\e{}_{\s}{}_{\m}{}_{\n}{}_{\rho} \, \psi_b  
 ~~~,
} \label{QTd2}
\ee 
\noindent
$ {Hodge-Dual~ \#3~Chiral~Supermultiplet: ~(A, \, B, \,  \psi_a , \, 
{\rm f}_{\m \, \n \, \rho}, \, {\rm g}_{\m \, \n \, \rho})}$
\be
 \eqalign{
{~~~~} {\rm D}_a A ~&=~ \psi_a  ~~~~~~~~~~~~~~,~~~~
{\rm D}_a B ~=~ i \, (\g^5){}_a{}^b \, \psi_b  ~~~~~~~~~, \cr
{\rm D}_a \psi_b ~&=~ i\, (\g^\m){}_{a \,b}\,  \pa_\m A 
~-~  (\g^5\g^\m){}_{a \,b} \, \pa_\m B \cr
&~~~~-~ i \,  
\frac 1{3!} \, C_{a\, b} \, (\e{}^{\s}{}^{\m}{}^{\n}{}^{\rho} \, 
\pa_{\s} {\rm f}_{\m \, \n \, \rho})
 ~+~  \frac 1{3!} \, (\g^5){}_{ a \, b}
\, (\e{}^{\s}{}^{\m}{}^{\n}{}^{\rho} \, \pa_{\s} {\rm g}_{\m
 \, \n \, \rho})   ~~, \cr
{\rm D}_a {\rm f}_{\m \, \n \, \rho} ~&=~ -\,  (\g^\s){}_a{}^b \,  
\e{}_{\s}{}_{\m}{}_{\n}{}_{\rho} \, \psi_b    
~~~,~~~ 
{\rm D}_a {\rm g}_{\m \, \n \, \rho} ~=~ -\,  (\g^5 \g^\s){}_a{}^b \,  
\e{}_{\s}{}_{\m}{}_{\n}{}_{\rho} \, \psi_b    ~~~.
} \label{QTd3}
\ee  

For the $ {Hodge-Dual~ \#1~Chiral~Supermultiplet}$ one should perform
the replacement of the auxiliary fields according to
\be
\int \, dt \, F ~\to~ {\rm f}_{ 1 2 3}  ~~~,~~~ G ~\to~ G  ~~~, 
\label{R_1}
\ee
where f$ {}_{ 1 2 3}$ is the purely spatial component of the Lorentz
3-form  ${\rm f}_{\m\n\r}$.

For the $ {Hodge-Dual~ \#2~Chiral~Supermultiplet}$ one should perform
the replacement of the auxiliary fields according to
\be
F ~\to~ F   ~~~,~~~ 
\int \, dt \, G ~\to~ {\rm g}_{ 1 2 3}  
~~~, 
\label{R_2}
\ee
where g$ {}_{ 1 2 3}$ is the purely spatial component of the Lorentz
3-form  ${\rm g}_{ \m\n\r}$.

For the $ {Hodge-Dual~ \#3~Chiral~Supermultiplet}$ one should perform
the replacement of the auxiliary fields according to
\be
\int \, dt \, F ~\to~ {\rm f}_{ 1 2 3}  ~~~,~~~
\int \, dt \, G ~\to~ {\rm g}_{ 1 2 3}  
~~~, 
\label{R_3}
\ee
where f$ {}_{ 1 2 3}$ is the purely spatial component of the Lorentz 3-form  ${\rm f
}_{ \m\n\r}$ and g$ {}_{ 1 2 3}$ is the purely spatial component of the Lorentz
3-form  ${\rm g}_{ \m\n\r}$.

$ {Vector~Supermultiplet:~ (A{}_{\mu} , \, \l_b , \,  {\rm d})}$
\be
\eqalign{
{~~~~} {\rm D}_a \, A{}_{\mu} ~&=~  (\gamma_\mu){}_a {}^b \,  \l_b  ~~~, \cr
{\rm D}_a \l_b ~&=~   - \,i \, \fracm 14 ( [\, \gamma^{\mu}\, , \,  \gamma^{\nu} 
\,]){}_a{}_b \, (\,  \pa_\mu  \, A{}_{\nu}    ~-~  \pa_\nu \, A{}_{\mu}  \, )
~+~  (\gamma^5){}_{a \,b} \,    {\rm d} ~~,  {~~~~~~~}  \cr
{\rm D}_a \, {\rm d} ~&=~  i \, (\gamma^5\gamma^\mu){}_a {}^b \, 
\,  \pa_\mu \l_b  ~~~, \cr
}  \label{QT2}
\ee  \vskip 0.12in \noindent
${Axial-Vector~Supermultiplet:~ (U{}_{\mu} , \, {\Tilde \l}_b , \,  {\Tilde {\rm d}}
)}$
\be
\eqalign{
{~~~~} {\rm D}_a \, U{}_{\mu} ~&=~ i\, (\gamma^5 \gamma_\mu){}_a {}^b 
\,  {\Tilde \l}_b  ~~~, \cr
{\rm D}_a {\Tilde \l}_b ~&=~  \, \fracm 14 ( \gamma^5 [\, \gamma^{\mu}\, , \,  
\gamma^{\nu} \,]){}_a{}_b \, (\,  \pa_\mu  \, U{}_{\nu}    ~-~  \pa_\nu 
\, U{}_{\mu}  \, ) ~+~ i \, C{}_{a \,b} \, {\Tilde {\rm d}} ~~,  {~~~~~~~}  \cr
{\rm D}_a \,  {\Tilde {\rm d}} ~&=~  - \, (\gamma^\mu){}_a {}^b \, 
\,  \pa_\mu {\Tilde \l}_b  ~~~, 
\cr}  \label{QT2a}
\ee
${Hodge-Dual~Vector~Supermultiplet:~ (A{}_{\mu} , \, \l_b , \,  {\rm d}
{}_{\mu \, \nu \, \rho} 
)}$
\be
\eqalign{
{~~~~~~~~~~~~~~~~} {\rm D}_a \, A{}_{\mu} ~&=~  (\gamma_\mu){}_a 
{}^b \,  \l_b  ~~~, \cr
{\rm D}_a \l_b ~&=~   - \,i \, \fracm 14 ( [\, \gamma^{\mu}\, , \, \gamma^{\nu} 
\,]){}_a{}_b \, (\,  \pa_\mu  \, A{}_{\nu}    ~-~  \pa_\nu \, A{}_{\mu}  \, )
~+~  \frac 1{3!} \, (\gamma^5){}_{ a \, b} \, (\epsilon{}^{\s}{}^{\mu}{}^{\nu}
{}^{\rho} \, \pa_{\s} {\rm d}{}_{\mu \, \nu \, \rho} )  ~~,  {~~~~~~~}  \cr
{\rm D}_a \, {\rm d}_{\m\n\rho} ~&=~  -i \, (\gamma^5\gamma^\s){}_a {}^b \, 
\,  \e_{\s\m\n\rho} \, \l_b  ~~~, \cr
}  \label{QTd2a}
\ee 
\noindent
${Hodge-Dual~ Axial-Vector~Supermultiplet:~ (A{}_{\mu} , \, {\Tilde \l}
{}_b , \,  {\Tilde {\rm d}}{}_{\mu \, \nu \, \rho} )}$
\be
\eqalign{
{~~~~} {\rm D}_a \, U{}_{\mu} ~&=~ i\, (\gamma^5 \gamma_\mu){}_a {}^b 
\,  {\Tilde \l}_b  ~~~, \cr
{\rm D}_a {\Tilde \l}_b ~&=~  \, \fracm 14 ( \gamma^5 [\, \gamma^{\mu}\, , \,  
\gamma^{\nu} \,]){}_a{}_b \, (\,  \pa_\mu  \, U{}_{\nu}    ~-~  \pa_\nu 
\, U{}_{\mu}  \, ) ~+~ i \,  \frac 1{3!} \, C{}_{a \,b} \, \, (\epsilon{}^{\s}{}^{\mu}
{}^{\nu}{}^{\rho} \, \pa_{\s} {\Tilde {\rm d}}{}_{\mu \, \nu \, \rho} )  ~~,    \cr
{\rm D}_a \,  {\Tilde {\rm d}}_{\m\n\rho} ~&=~   (\gamma^\s){}_a {}^b \, \e_{\s\m\n\rho}\,
 {\Tilde \l}_b    ~~~, \cr
}  \label{QTd2b}
\ee

For $Axial-Vector~Supermultiplet$ one should perform
the replacement of the fermionic fields according to
\be
 \l_b ~\to~ -i\, (\g^5)_b{}^c \, \Tilde{\l}_c  ~~~,
\label{AV_1}
\ee
and the bosonic fields according to 
\be
A_{\mu} ~\to~  U{}_{\mu} ~~~,~~~{\rm d} ~\to~  {\Tilde {\rm d}} ~~~.
\label{AV_2}
\ee

For $Hodge-Dual~Vector~Supermultiplet$ one should perform the replacement of the the auxiliary field according to
\be
\int \, dt \, {\rm d} ~\to~ {\rm d}_{ 1 2 3}  ~~~,
\label{HV}
\ee
where $ {\rm d}_{ 1 2 3}$ is the purely spatial component of the Lorentz
3-form  ${\rm d}_{\m\n\r }$.
Note that the last line in Equation (3.9) in \cite{HYMN1} has typos and the corrected one is (\ref{QTd2a}). 

For $Hodge-Dual~ Axial-Vector~Supermultiplet$ one should perform the replacement of the auxiliary field according to 
\be
\int \, dt \, \Tilde{\rm d} ~\to~ \Tilde{\rm d}_{ 1 2 3}  ~~~,
\label{HAV}
\ee
where $ \Tilde{\rm d}_{ 1 2 3}$ is the purely spatial component of the Lorentz
3-form  $\Tilde{\rm d}_{\m\n\r }$.
Note that the last line in Equation (3.10) in \cite{HYMN1} has typos and the corrected one is (\ref{QTd2b}). 

$ {Tensor~Supermultiplet: ~(\varphi, \, B{}_{\mu \, \nu }, \,  \chi_a )}$
\be
 \eqalign{
{\rm D}_a \varphi ~&=~ \chi_a  ~~~,~~~~~ 
{\rm D}_a B{}_{\mu \, \nu } ~=~ -\, \fracm 14 ( [\, \gamma_{\mu}
\, , \,  \gamma_{\nu} \,]){}_a{}^b \, \chi_b  ~~~, \cr
{\rm D}_a \chi_b ~&=~ i\, (\gamma^\mu){}_{a \,b}\,  \pa_\mu \varphi 
~-~  (\gamma^5\gamma^\mu){}_{a \,b} \, \e{}_{\mu}{}^{\r \, \s \, \t}
\pa_\r B {}_{\s \, \t}~~, {~~~~~~~~~~~~~~\,~~}
}  \label{QT3}
\ee
 \noindent
$ {Axial-Tensor~Supermultiplet: ~( {\Tilde {\varphi}}, \, 
C{}_{\mu \, \nu }, \,  {\Tilde {\chi}} {}_a )}$
\be
 \eqalign{
{\rm D}_a {\Tilde {\varphi}} ~&=~  i \, (\gamma^5){}_{a}{}^{b} \, {\Tilde {\chi}}{}_b  ~~~,~~~~~ 
{\rm D}_a C{}_{\mu \, \nu } ~=~ -i\, \fracm 14  ( \gamma^5 [\, \gamma_{\mu}
\, , \,  \gamma_{\nu} \,]){}_a{}^b \,  {\Tilde {\chi}}{}_b  ~~~, \cr
{\rm D}_a  {\Tilde {\chi}}{}_b ~&=~ -\, (\gamma^5 \gamma^\mu){}_{a \,b}\,  \pa_\mu 
 {\Tilde {\varphi}}
~-~  i \, (\gamma^\mu){}_{a \,b} \, \e{}_{\mu}{}^{\r \, \s \, \t}
\pa_\r C{}_{\s \, \t}~~, {~~~~~~~~~~~~~~\,~~}
}  \label{QT3a}
\ee

For $Axial-Tensor~Supermultiplet$ one should perform the replacement of the fermionic fields according to 
\be
 \chi_a ~\to~ i\, (\g^5)_a{}^b \, \Tilde{\chi}_b  ~~~,
\label{AT_1}
\ee
and the bosonic fields according to 
\be
\varphi ~\to~  \Tilde{\varphi} ~~~,~~~B{}_{\mu \, \nu } ~\to~  C{}_{\mu \, \nu } ~~~.
\label{AT_2}
\ee
Note that the first line in Equation (3.12) in \cite{HYMN1} has typos and the corrected one is (\ref{QT3a}). 
  
\newpage
\section{Traces and Determinants}\label{a:EigenSumProd}
As mentioned in the body of the paper, the traces and determinants of the various Banchoff matrices presented are an important calculation as to the classification of the HYMNs. These are listed below.

\begin{table}[htp!]
\centering
\begin{tabular}{|c|c|c|c|c|}
\hline
& ${\bm B}_L$ & ${\bm B}_R$ & $|{\bm B}_L|$ & $|{\bm B}_R|$ \\
\hline
 trace & $4 \rho_B$ & $-2 \, (\,1 \,+\,  \rho_B^2 \,) $ & $4 \rho_B$ & $2 \, (\,1 \,+\,  \rho_B^2 \,)  $ \\
\hline
determinant & $\rho_B^4$ & $\rho_B^4$ & $\rho_B^4$ &  $\rho_B^4$ \\
\hline
\end{tabular}
\caption{Results of $\text{trace}$ and determinant eigenvalues with/without 
dashings for the adinkra in Fig \ref{fig:CMadink}}
\label{tab:result_GR(4,4)CMa}
\end{table}

\begin{table}[htp!]
\centering
\begin{tabular}{|c|c|c|c|c|}
\hline
& ${\bm B}_L$ & ${\bm B}_R$ & $|{\bm B}_L|$ & $|{\bm B}_R|$ \\
\hline
 trace & $-2 \, (\,1 \,+\,  \rho_B^2 \,)$ & $2 \, (\,1 \,+\,  \rho_B^2 \,) $ & $2 \, (\,1 \,+\,  \rho_B^2 \,) $ & $ 2 \, (\,1 \,+\,  \rho_B^2 \,) $ \\
\hline
determinant & $\rho_B^2$ & $\rho_B^2$ & $\rho_B^2$ &  $\rho_B^2$ \\
\hline
\end{tabular}
\caption{Traces and determinants with/without 
dashings for the adinkra in Fig \ref{fig:VMadink}}
\label{tab:result_GR(4,4)VMa}
\end{table}

\begin{table}[htp!]
\centering
\begin{tabular}{|c|c|c|c|c|}
\hline
& ${\bm B}_L$ & ${\bm B}_R$ & $|{\bm B}_L|$ & $|{\bm B}_R|$ \\
\hline
 trace & $-4$ & $4$ & $4$ & $4$ \\
\hline
determinant & $1$ & $1$ & $1$ &  $1$ \\
\hline
\end{tabular}
\caption{Traces and determinants with/without 
dashings for the adinkra in Fig \ref{fig:TMadink}}
\label{tab:result_GR(4,4)TMa}
\end{table}

\begin{table}[htp!]
\centering
\begin{tabular}{|c|c|c|c|c|}
\hline
& ${\bm B}_L$ & ${\bm B}_R$ & $|{\bm B}_L|$ & $|{\bm B}_R|$ \\
\hline
 trace & $4 \chi_{\rm o}$ & $-4 \chi_{\rm o}$ & $4$ & $4$ \\
\hline
determinant & $1$ & $1$ & $1$ &  $1$ \\
\hline
\end{tabular}
\caption{Traces and determinants with/without 
dashings for the 36, 864 valise adinkras \newline $~~~~~~~\,~~~~~~~\,$
 associated with (4, 0) SUSY}
\label{tab:result_GR(4,4)xa}
\end{table}

where 
\begin{equation}
    \chi_o({\mathcal R}) ~=~ \begin{cases}
    1 ~~~~~\text{when }\mathcal{R}=\text{SM-I, SM-IV}\\
    -1 ~~~\text{when }\mathcal{R}=\text{SM-II, SM-III}\\
    \end{cases}
\end{equation}
which is true for all the valise GR(4,4) adinkras.

\begin{table}[htp!]
\centering
\begin{tabular}{|c|c|c|c|c|}
\hline
& ${\bm B}_L$ & ${\bm B}_R$ & $|{\bm B}_L|$ & $|{\bm B}_R|$ \\
\hline
 trace & $2 \, (\,1 \,+\,  \rho_B^2 \,) $ & $-2 \, (\,1 \,+\,  \rho_B^2 \,) $ & $2 \, (\,1 \,+\,  \rho_B^2 \,) $ & $2 \, (\,1 \,+\,  \rho_B^2 \,) $ \\
\hline
determinant & $\rho_B^2$ & $\rho_B^2$ & $\rho_B^2$ &  $\rho_B^2$ \\
\hline
\end{tabular}
\caption{Traces and determinants with/without 
dashings for Hodge-Dual \#1 and Hodge- \newline \indent $~~~~~~~~~~~~~~$
Dual \#2 Chiral Supermultiplet}
\label{tab:result_HodgeDual-1CMa}
\end{table}

\begin{table}[htp!]
\centering
\begin{tabular}{|c|c|c|c|c|}
\hline
& ${\bm B}_L$ & ${\bm B}_R$ & $|{\bm B}_L|$ & $|{\bm B}_R|$ \\
\hline
 trace & $4$ & $-4$ & $4$ & $4$ \\
\hline
determinant & $1$ & $1$ & $1$ &  $1$ \\
\hline
\end{tabular}
\caption{Traces and determinants with/without 
dashings for Hodge-Dual \#3 Chiral \newline $~~~~~~~~~~~~~~~$
Supermultiplet}
\label{tab:result_HodgeDual-3CMa}
\end{table}

\begin{table}[htp!]
\centering
\begin{tabular}{|c|c|c|c|c|}
\hline
& ${\bm B}_L$ & ${\bm B}_R$ & $|{\bm B}_L|$ & $|{\bm B}_R|$ \\
\hline
 trace & $-2 \, (\,1 \,+\,  \rho_B^2 \,) $ & $2 \, (\,1 \,+\,  \rho_B^2 \,) $ & $2 \, (\,1 \,+\,  \rho_B^2 \,) $ & $2 \, (\,1 \,+\,  \rho_B^2 \,) $ \\
\hline
determinant & $\rho_B^2$ & $\rho_B^2$ & $\rho_B^2$ &  $\rho_B^2$ \\
\hline
\end{tabular}
\caption{Traces and determinants with/without 
dashings for Axial-Vector Supermultiplet}
\label{tab:result_Axial-VMa}
\end{table}

\begin{table}[htp!]
\centering
\begin{tabular}{|c|c|c|c|c|}
\hline
& ${\bm B}_L$ & ${\bm B}_R$ & $|{\bm B}_L|$ & $|{\bm B}_R|$ \\
\hline
 trace & $-4$ & $4$ & $4$ & $4$ \\
\hline
determinant & $1$ & $1$ & $1$ &  $1$ \\
\hline
\end{tabular}
\caption{Traces and determinants with/without 
dashings for Hodge-Dual Vector Super- \newline $~~~~~~\,~~~~~~\,~~$
multiplet, Hodge-Dual Axial-Vector Supermultiplet, and Axial-Tensor Supermultiplet}
\label{tab:result_HodgeDual-VMa}
\end{table}

For the adinkra described in Chapter \ref{sec:4colornonmininal}, since as we already seen that eigenvalues are not sensitive to dashings, we will only show ${\rm Tr}(\bm B_L)$, ${\rm Tr}(\bm B_R)$, ${\rm Det}(\bm B_L)$, and ${\rm Det}(\bm B_R)$ in Table \ref{tab:Function_GR84}.

\begin{table}[htp!]
\centering
\begin{tabular}{|c|c|c|c|c|}
\hline
& ${\rm Tr}(\bm B_L)$ & ${\rm Tr}(\bm B_R)$ & ${\rm Det}(\bm B_L)$ & ${\rm Det}(\bm B_R)$ \\
\hline\hline
$<8|8>$ & 0 & 0 & 1 & 1\\\hline
$<7|8|1>$ & 0 & 0 & $\rho_B^2$ & $\rho_B^2$ \\\hline
$<6|8|2>$ & 0 & 0 & $\rho_B^4$ & $\rho_B^4$ \\\hline
$<5|8|3>$ & 0 & 0 & $\rho_B^6$ & $\rho_B^6$ \\\hline
$<4|8|4>$ & 0 & 0 & $\rho_B^8$ & $\rho_B^8$ \\\hline
$<3|8|5>$ & 0 & 0 & $\rho_B^{10}$ & $\rho_B^{10}$ \\\hline
$<2|8|6>$ & 0 & 0 & $\rho_B^{12}$ & $\rho_B^{12}$ \\\hline
$<1|8|7>$ & 0 & 0 & $\rho_B^{14}$ & $\rho_B^{14}$ \\\hline
$<1|7|7|1>$ & 0 & 0 & $\rho_B^{14}\rho_F^2$ & $\rho_B^{14}\rho_F^2$  \\\hline
$<1|6|7|2>$ & 0 & 0 & $\rho_B^{14}\rho_F^4$ & $\rho_B^{14}\rho_F^4$  \\\hline
$<1|5|7|3>$ & 0 & 0 & $\rho_B^{14}\rho_F^6$ & $\rho_B^{14}\rho_F^6$  \\\hline
$<1|4|7|4>$ & 0 & 0 & $\rho_B^{14}\rho_F^8$ & $\rho_B^{14}\rho_F^8$  \\\hline
$<1|4|6|4|1>$ & 0 & 0 & $\rho_B^{16}\rho_F^8$ & $\rho_B^{16}\rho_F^8$  \\\hline
\end{tabular}
\caption{Traces and determinants for adinkras described in Chapter \ref{sec:4colornonmininal}}
\label{tab:Function_GR84}
\end{table}
In summary, for the 4-color nonminimal adinkras described in Chapter \ref{sec:4colornonmininal}, we have general equations
\begin{equation}
    \begin{split}
        &{\rm Tr}(\bm B_L) ~=~ {\rm Tr}(\bm B_R) ~=~ 0\\
        &{\rm Det}(\bm B_L) ~=~ {\rm Det}(\bm B_R) ~=~ \rho_B^{2\times\text{\# of bosons lifted}}\rho_F^{2\times\text{\# of fermions lifted}}
    \end{split}
\end{equation}

For two five-color adinkras described in Chapter \ref{sec:2-5CLR}, since eigenvalues are not sensitive to dashings as well, we will only show results with dashings. 
\begin{table}[htp!]
\centering
\begin{tabular}{|c|c|c|}
\hline
1st Adinkra & ${\bm B}_L{\bm B}_R$ & ${\bm B}_R{\bm B}_L$  \\
\hline
 trace & $2(\, 1 + \rho_B + 2\rho_B^2 \,)$ & $2(\, 1 + \rho_B + 2\rho_B^2 \,)$ \\
\hline
determinant & $\rho_B^{10}$ & $\rho_B^{10}$  \\
\hline \hline
2nd Adinkra & ${\bm B}_L{\bm B}_R$ & ${\bm B}_R{\bm B}_L$  \\
\hline
 trace & $2(\,3\rho_B + \rho_B^2 \,)$ & $2(\,3\rho_B + \rho_B^2 \,)$ \\
\hline
determinant & $\rho_B^{10}$ & $\rho_B^{10}$  \\
\hline 
\end{tabular}
\caption{Traces and determinants for the adinkras in Fig \ref{fig:N5adink}}
\label{tab:result_GR(8,5)}
\end{table}

For two six-color adinkras described in Chapter \ref{sec:2-6color}, we have Table \ref{tab:result_GR(8,6)}. 
\begin{table}[htp!]
\centering
\begin{tabular}{|c|c|c|c|c|}
\hline
1st Adinkra & ${\bm B}_L$ & ${\bm B}_R$ & $|{\bm B}_L|$ & $|{\bm B}_R|$ \\
\hline
 trace & 0 & 0 & 0 & 0 \\
\hline
determinant & $\rho_B^6$ & $\rho_B^6$ & $\rho_B^6$ &  $\rho_B^6$ \\
\hline \hline
2nd Adinkra & ${\bm B}_L$ & ${\bm B}_R$ & $|{\bm B}_L|$ & $|{\bm B}_R|$ \\
\hline
 trace &  0 & 0 & 0 & 0 \\
\hline
determinant  & $\rho_B^6$ & $\rho_B^6$ & $\rho_B^6$ &  $\rho_B^6$ \\
\hline
\end{tabular}
\caption{Traces and determinants with/without 
dashings for the adinkras in Fig \ref{fig:N6adink}}
\label{tab:result_GR(8,6)}
\end{table}

\newpage

\end{document}